\documentclass{aa}
\usepackage{graphicx}
\begin{document}

\thesaurus{06(08.14.1; 08.18.1; 08.16.6; 02.07.1; 02.18.8; 03.13.4)}

\title{Construction of highly accurate models of rotating neutron
stars  --  comparison of three different numerical schemes }
\titlerunning{Highly accurate models of rotating neutron stars}

\author{T. Nozawa\inst{1}
\and N. Stergioulas\inst{2}
\thanks{\emph{Present adress:}
 Max Planck Institute for Gravitational Physics,
The Albert Einstein Institute,
D-14473 Potsdam, Germany}
\and E. Gourgoulhon\inst{3}
\and Y. Eriguchi\inst{1}}

\institute{Department of Earth Science and Astronomy, Graduate School
of Arts and Sciences, University of Tokyo, Komaba, Meguro, Tokyo 153,
Japan
\and Department of Physics, University of Wisconsin-Milwaukee, 
PO box 413, Milwaukee, WI 53201, USA
\and D{\'e}partement d'Astrophysique Relativiste et de Cosmologie, 
 UPR 176 du CNRS, 
Observatoire de Paris, F-92195 Meudon Cedex, France}

\offprints{E.~Gourgoulhon}

\date{Received / Accepted}

\maketitle

\begin{abstract}
We conduct a direct comparison of three different representative
numerical codes for constructing models of rapidly rotating neutron
stars in general relativity. Our aim is to evaluate the accuracy of
the codes and to investigate how the accuracy is affected by the
choice of interpolation, domain of integration and equation of
state. In all three codes, the same physical parameters, 
equations of state and interpolation method are used. We
construct 25 selected models for polytropic equations of state and 22
models with realistic neutron star matter equations of state. 
The three codes agree  well with each other (typical agreement
is better than 0.1 \% to 0.01 \%) for most models, except for the extreme
assumption of uniform density stars.  We conclude that the codes can
be used for the construction of highly accurate initial data configurations
for polytropes of index $N>0.5$ (which typically correspond to realistic 
neutron stars), when the domain of integration includes
all space and for realistic equations with no phase transitions. 
With the exception of the uniform density case, the obtained values of physical parameters 
for the models considered in this paper can be regarded as ``standard'' and we display them
in detail for all models.
\end{abstract}

\thesaurus{06(08.14.1; 08.18.1; 08.16.6; 02.07.1; 02.18.8; 03.13.4)}


\section{Introduction}

The physical state of the neutron star matter has not been fully
understood yet because it is very difficult to investigate particle
interactions beyond nuclear matter density ($\varepsilon_{\rm N}/c^2
\sim 2 \times 10^{14}$ g cm$^{-3}$) either from nuclear experiments or
from nuclear theories, (here $\varepsilon_{\rm N}$ is the energy
density of the nuclear matter and $c$ is the velocity of light).
%%Throughout this paper we use the geometrized units, i.e.  $c = G = 1$. 
Given this situation, one promising approach to explore the behavior
of very high density matter is to make use of the macroscopic
quantities of neutron stars. In particular, the mass and the
rotational period of neutron stars depend crucially on the softness of
the equation of state (EOS) at very high densities (see e.g. Friedman et
al. 1984, 1986, 1989), thus, observational constrains, matched with
theoretical models, may help in reconstructing the equation of state of
very high density matter.
 
Given a particular equation of state, the mass of neutron stars varies
with central energy density and always reaches a maximum. This implies
that if the maximum mass of neutron star models constructed with a
certain equation of state is smaller than the mass of observed neutron
stars, that equation of state must be discarded. Currently, the largest
accurately measured mass of slowly rotating neutron stars is $M_{\rm
BP} = 1.44 M_{\odot}$, where $M_{\rm BP}$ is the mass of one of
the components of the binary pulsar PSR1913+16 (Taylor \& Weisberg 
\cite{tayl89}) and
$M_{\odot}$ is one solar mass.  Individual masses of
neutron stars have also been estimated in six other binary pulsars
(Thorsett et al. \cite{thor93}, Wolszczan \cite{wol97}), as well as in six X-ray binaries
(van Kerkwijk et al. \cite{vankerk95}) but the accuracy is not as good as in PSR1913+16.
Thus, equations of state which give larger masses than $M_{\rm BP}$
for slowly rotating stars, can be valid as candidates for the
real equation of state at very high densities. Since the maximum mass of
neutron stars is smaller for more compressible (soft) equations
of state than for less compressible (stiff) equations of state, the
true equation of state at high densities cannot be extremely soft.

On the other hand, stiff equations of state can be limited by
considering the neutron star with the shortest rotational period,
i.e. the most rapidly rotating pulsar.  There exists a lower limit on
the rotational period for each equation of state, because if the
centrifugal force exceeds the self-gravity at the equatorial surface,
no equilibrium states are allowed. The lower limit of the rotational
period depends on the softness of the equation of state - the radius
of neutron stars with softer equations of state is smaller, which
allows for higher rotation rates. Thus, if very rapidly rotating
neutron stars should be found, we could exclude most stiff equations
of state. At the moment, the shortest period of observed pulsars is
1.56 ms, of PSR1937+21. Consequently, equations of state for which the
shortest rotational period is larger than this value, must be excluded
as candidates for the real equation of state for neutron star matter.

The discussions above require us to make use of highly accurate
schemes for constructing rotating neutron star models, in order to
compute precise theoretical values of masses and rotational periods.
Highly accurate relativistic equilibrium models are also needed as
initial data for relativistic time-evolution codes (modeling of
nonlinear pulsations, collapse and generation of gravitational waves).
Recently, a number of groups have succeeded in constructing models of
rapidly rotating neutron stars (Friedman et al. 1984, 1986, 1988,
1989, Eriguchi et al. \cite{erig94}, Salgado et al. \cite{salg94}, \cite{salg94b}, 
Cook et al. \cite{cook94b},
Stergioulas \& Friedman \cite{stag95} -- for a recent review see
Stergioulas \cite{S98}).  However, the obtained models by those authors do not
always agree with each other (see e.g. Friedman et al. \cite{frie89}, 
Eriguchi et al. \cite{erig94},
Salgado et al. \cite{salg94}, Cook et al. \cite{cook94b}, Stergioulas \& Friedman 
\cite{stag95}). Although Stergioulas \&
Friedman~(1995) have determined the cause of the discrepancy between
models in Friedman et al.~(1989) and Eriguchi et al.~(1994), (which
was due to the use of a slightly different equation of state table),
the reasons of smaller differences which remain, even after using
exactly the same equation of state, have not been clarified yet. This
is because numerical techniques used in the different codes, such as
the choice of parameters defining the model, the interpolation method,
the method of integrating the field equations, a.s.o. are not the
same.

In this paper, three groups using their own codes (Komatsu et al.
\cite{koma89a}, Eriguchi et al. \cite{erig94}, Salgado et al.
\cite{salg94}, Stergioulas \& Friedman \cite{stag95}) will decrease
the differences between their results to a minimum possible, by tuning
each code and using the same parameters, the same schemes of
interpolation, the same equations of state, and so on.  Since the
basic schemes used by the three groups are different, it will be
impossible to have exactly the same results and the relative
differences between results are a measure of the accuracy of the
codes.  Models obtained with small relative differences between the
three codes can be considered as ``standard" models for each equation
of state.  Furthermore, this direct comparison allows us to
investigate the effect that the choice of interpolation method,
equation of state and domain of integration has on the accuracy of the
codes.

\section{Construction of neutron star models}

If rapidly rotating neutron stars were nonaxisymmetric, 
they would emit gravitational waves in a very short time scale and 
settle down to axisymmetric configurations. Moreover, the gravity of 
typical neutron star is strong because 
\begin{equation}
  \frac{2 G M_{\rm ns}}{c^2 R_{\rm ns}} \sim 0.4,  
\end{equation}
where $G$ is the gravitational constant, and $M_{\rm ns} \sim 1.4 M_{\sun}$ 
and $R_{\rm ns} \sim 10 {\rm km}$ are the mass and the radius of a typical 
neutron star. Therefore, we need to solve for rotating and axisymmetric 
configurations in the framework of general relativity.  

For the matter and the spacetime we make the following assumptions:
\begin{enumerate}
\item  The matter distribution and the spacetime are axisymmetric.
\item  The matter and the spacetime are in a stationary state.
\item  The matter has no meridional motions. The only motion of 
the matter is a circular one that is represented by the angular velocity.
\item  The angular velocity $\Omega$ is constant, as seen by a distant observer at rest. 
\item  The matter can be described as a perfect fluid. 
\end{enumerate}

Under these assumptions, the metric can be expressed as follows
(Papapetrou 1966, Carter 1969):
%
%\begin{mathletters}
\begin{eqnarray}
 ds^2 & = & -  e^{2\nu} dt^2 + e^{2\alpha}(dr^2 + r^2 d \theta^2)  \nonumber \\ && 
+ e^{2\beta} r^2 \sin^2 \theta (d \varphi - \omega dt)^2,  \\
           & = & - e^{2\nu} dt^2 + e^{2\zeta-2\nu}( dr^2 + r^2 d \theta^2) \nonumber \\ &&  
+ B^2 e^{-2\nu} r^2 \sin^2 \theta (d \varphi - \omega dt)^2,  \\
           & = & - e^{2\nu} dt^2 + e^{2\mu}( dr^2 + r^2 d \theta^2) 
+ e^{2\psi} (d \varphi - \omega dt)^2, 
\end{eqnarray}
\label{metric}
%\end{mathletters}
%
\noindent
where $t$ is the time coordinate and the polar coordinates $(r, \theta, 
\varphi)$ are used. The metric depends on four metric functions (or
metric potentials) which are functions of $r$ and $\theta$ only. Different
authors have used the set of functions $(\nu, \omega, \alpha, \beta)$,
$(\nu, \omega, \zeta, B)$ or $(\nu, \omega, \mu, \psi)$, which are related
to each other through (\ref{metric}). In (\ref{metric}) and throughout the
text we use gravitational units ($c=G=1$), unless otherwise stated.

The energy momentum tensor $T^{ab}$ is expressed as
\begin{equation}
T^{ab} = (\varepsilon + p) u^au^b + p g^{ab},  
\end{equation}
where $\varepsilon$, $p$ and $u^a$ are the energy density, the pressure 
and the four-velocity, respectively. In the coordinate basis defined
by (\ref{metric}), the components of the four-velocity are
\begin{equation}
 u^a = \frac{e^{-\nu}}{\sqrt{1 - v^2}} (1, 0, 0, \Omega),
\end{equation}
where the proper velocity $v$ with respect to a local zero angular momentum 
observer is defined by %
\begin{equation}
  v \equiv r \sin \theta e^{\beta- \nu} (\Omega - \omega).
\end{equation}

Using the metric and the energy momentum tensor mentioned above,
we can write down the Einstein equations for the metric components.
Although we omit detailed expressions for the Einstein equations here, 
one can easily derive them by straightforward calculations or consult 
the papers by Butterworth \& Ipser~(1976), Komatsu et al.~(1989a)
and Bonazzola et al.~(1993).

The equation of hydrostationary equilibrium can be derived from the 
equations of motion and takes the following form:
\begin{equation}
\frac{1}{\varepsilon + p} \nabla p + \nabla \nu - \frac{1}{2} 
\nabla \ln(1-v^2) = 0.
\label{hydrost}
\end{equation}
This equation can be integrated, if we specify the equation of state
which relates the energy density to the pressure. For a given EOS, a
model is uniquely specified by fixing two parameters, such as the
central energy density $\epsilon_c$ and the ratio of the polar to the
equatorial coordinate radius, $r_p/r_e$. Then, four Einstein field
equations and the equation of hydrostationary equilibrium must be
solved with appropriate boundary conditions, to yield the four metric
functions and the density distribution. The available codes for
obtaining relativistic rotating neutron star models differ basically
in the choice and method of integration of the four field equations
and in the finite grid and finite difference scheme used for the
integration.

\section{Codes}

We will compare three different codes, which follow the  KEH scheme, 
developed by Komatsu, Eriguchi and 
Hachisu~(1989a, 1989b) for general relativistic polytropes with uniform 
and differential rotation and improved by Cook, Shapiro \& Teukolsky~(1992, 1994a,b)
and the BGSM scheme due to Bonazzola, Gourgoulhon, Salgado \&  Marck~(1993). 

Concering the KEH scheme, for the present comparison, we will  use the 
original KEH code, KEH(OR),  and the KEH code by Stergioulas \& 
Friedman~(1995), KEH(SF), which follows the Cook, Shapiro \& Teukolsky~(1992, 
1994a,b) modification of the KEH scheme.

\subsection{A short description of each code}

In this section the three different numerical codes are briefly described. 
Details of these codes can be found in  Komatsu et 
al.~(1989a, 1989b) and in Eriguchi et al.~(1994) for the KEH(OR) code, 
in Stergioulas \& Friedman~(1995) for the KEH(SF) code and in 
Bonazzola et al.~(1993) for the BGSM code.

\subsubsection{The KEH(OR) code}

Komatsu et al.~(1989a) have developed a new scheme for solving rapidly 
rotating relativistic stars. The Einstein equations for three metric 
potentials $\nu, \beta$ and $\omega$ are transformed into integral equations
by using appropriate Green's functions for the elliptical type differential
operators. In principle, one can choose Green's functions which decrease
as $1/r$ or more rapidly at large distances.  Consequently boundary conditions 
at infinity, i.e. asymptotically flat conditions can be easily included 
in the integral equations.  It is noted that in this integral representation
the integrand contains the metric and the matter quantities such as
the energy density and the pressure. The fourth metric $\alpha$ obeys a 
first order partial differential equation which can be easily 
integrated, if the other metric potentials are known. The domain of integration
is truncated at a finite distance from the star (roughly twice the equatorial radius)
and the metric potentials 
%%%%%%%%%%%%%%%%%%%
in the integrands
%%%%%%%%%%%%%%%%%%%
are assumed to vanish at that finite distance (instead
of at infinity).

The KEH scheme is the extended version of the self-consistent-field
(SCF) scheme which was developed for solving Newtonian rotating stars
(Ostriker \& Mark \cite{ostr68}, Hachisu \cite{hach86}) and applied to
relativistic rotating stars by Bonazzola \& Schneider (1974) with a
choice of metric functions different from that of the codes considered
in this article. In the SCF method, the iteration proceeds as follows.
If one assumes initial guesses for the matter quantities and the
metric potentials, new (and better) values for $\nu, \beta$ and
$\omega$ can be obtained using the integral representations for the
metric potentials The fourth metric potential $\alpha$ can be easily
solved as mentioned before.  By using newly obtained metric
potentials, a new density and a new pressure can be computed from the
hydrostationary equilibrium equation (\ref{hydrost}).  One needs to
repeat the same procedure until the relative differences between the
newly obtained quantities and the old ones become less than a certain
small number, typically $10^{-5}$.

In the original KEH code, the ratio of the central pressure ($p_c$) 
to the central energy density ($\varepsilon_c$), 
\begin{equation}
\kappa \equiv \frac{p_c} {\varepsilon_c},
\end{equation}
and the ratio of the polar radius ($r_{\rm p}$) to the equatorial
radius ($r_{\rm e}$), $r_{\rm p}/r_{\rm e}$, are chosen as two
independent model parameters which specify the model uniquely.  The
KEH(OR) code used in this comparison differs from the code used in
Komatsu et al. (1989a) only in one aspect, that is mentioned in the
next section.

\subsubsection{The KEH(SF) code}
 
The KEH(SF) code (Stergioulas \& Friedman, 1995) differs from the
original KEH scheme in two ways. First, it follows Cook et al. 
(1992) in using a redefined radial variable
\begin{equation}
s \equiv \frac{r} {r + r_e}.
\end{equation}
where $r_e$ is the value of the coordinate $r$ at the equator.  By
this transformation, the region $[0, \infty]$ in the $r$ coordinate is
transformed to the region $[0, 1]$ in the $s$ coordinate.
Consequently, the domain of integration of the field equations does
not have to be truncated at a finite distance from the star (as in the
KEH(OR) code) but covers all space. With this choice, the boundary
conditions 
%%%%%%%%%%%%%
%%become 
%%%%%%%%%%%%%
can be applied accurately at infinity.

Second, Stergioulas \& Friedman found that the choice of coordinates
in the original KEH scheme results in the metric potential $\alpha$
oscillating in the radial direction.  The oscillation is especially
pronounced inside the star and introduces an error of 1-2 \% in the
mass, radius and other quantities. The problem was fixed by using a
finite difference formula for the second order radial derivative that
uses twice the grid-spacing. Although this formula is, in principle, of
lower accuracy, the oscillations are damped completely, resulting in 
a more accurate stellar model.
 
The KEH(OR) code used in this comparison has been modified so as to
use the same second order derivative formula as Stergioulas \&
Friedman, to smooth out the oscillations in the metric potential
$\alpha$.

\subsubsection{The BGSM code}

Bonazzola et al.~(1993) have developed a new formulation based on the 
$3 + 1$ formalism which has been used in hydrodynamics in general relativity.
Their choice of slicing and gauge in the $3 + 1$ formalism results in
the same form of the metric usually chosen for stationary problems, 
i.e. (\ref{metric}). Consequently the Einstein equations are
reduced to the same differential equations of elliptic type
as used by other schemes.

The main part of the BGSM formulation is similar to that of the KEH 
formulation, except for the metric coefficient $\zeta = \alpha + \nu$ for
which a second-order (elliptic) equation is used instead of a
first-order equation in KEH. The Einstein equations are reorganized 
so as to ``pick" out the Laplacian operators in two and three dimensional 
flat spaces and regard all other remaining terms as ``source terms" in 
the Poisson-like equations in two and three dimensional flat spaces.
Concerning the matter, essentially the same equation as (\ref{hydrost})
is used for the hydrostationary equation. 

The characteristic features of the BGSM code can be found in the numerical 
solving method, i.e. the pseudo-spectral method (Gottlieb \& Orszag 1977, 
Bonazzola et al. 1996, 1997, 1998b). In the spectral method 
all functions are expanded in terms of certain base functions and algebraic 
equations for coefficients which appear in the expansion are solved. 
Therefore there are two distinct procedures in this method: one is obtaining
coefficients from the functions and the other is constructing functions
by using the coefficients. Since these two steps need to be, in general, 
performed many times, it is highly desirable to use a fast algorithm.
In the spectral method of the BGSM code, Bonazzola et al.~(1993) have
adopted trigonometric functions for the angle variable and the Chebyshev
polynomials for the radial variable. Consequently for the angle part of 
any function, the fast fourier transform (FFT) can be employed. 
Concerning the radial variable, a new variable which is related to the
radial variable by a simple equation is introduced so that the Chebyshev
polynomials are expressed by the trigonometric functions.  After this
transformation, one can use the FFT also for the radial variable.  

The BGSM code can handle the region extended to infinity as is done by 
the KEH(SF) code. This can be performed by 
introducing a new radial variable $u$ as follows:
\begin{equation}
 u \equiv \frac{R_0} {r},  \qquad {\ \rm for \ outside \ of \ the \ matter}
\end{equation}
where $R_0$ is the maximum value of the stellar radius. 
The boundary conditions at infinity, are easily applied  
at $u = 0$.

It may be fair to note that in the Newtonian rotating star problems a similar 
expansion was used by Ostriker and Mark~(1968) in the SCF method, 
although Ostriker and Mark used the integral form of the Newtonian 
potential instead of solving the Poisson equation directly and they 
did not use the FFT.

\subsection{Relations among the three  different codes}

Here we summarize similarities and difference between the three codes:

A) Common features through all three codes:
\begin{enumerate}
\item Quasi-isotropic coordinates are used. It implies that the metric components
are essentially the same, although background views deriving the metric
are not the same. 
\item  Integral form of the hydrostationary equation for the matter is used.
\item  Poisson-like operators are ``picked" out from the Einstein equations
and the other terms are treated as ``source terms".
\end{enumerate}

B) The KEH(OR) code differs from the other two codes in that the
boundary conditions are not applied at infinity, but approximate
boundary conditions are applied at a finite distance from the star.

C) A difference between the BGSM code and the other codes is the use of a
second-order (elliptic)equation for $\zeta$ in BGSM versus a first-order equation for
$\alpha=\zeta-\nu$ in KEH.

We reorganized our codes so as to make the differences as small as possible. 
The codes agree exactly on:

\begin{enumerate}
\item the physical model parameters by which we can specify a model 
uniquely,
\item  the values of physical constants,  and
\item the equation of state of matter.
\end{enumerate}

Since the three codes use different grids and/or numerical methods for
solving the field equations, there will always be a residual
difference in the results, even after this reorganization. This
residual difference is what we want to determine.

\subsection{Starting point of computations}

\subsubsection{Constants}

Values used in this paper for the velocity of light $c$, 
the gravitational constant $G$, the mass of the sun $M_{\odot}$ and 
the baryon mass $m_{\rm B}$ are as follows:
\begin{eqnarray}
 c & = & 2.9979 \times 10^{10} {\rm cm \ s}^{-1}, \nonumber \\
 G & = & 6.6732 \times 10^{-8} {\rm g}^{-1} \ {\rm cm}^3 \  {\rm s}^{-2},
\nonumber \\
 M_{\odot} & = & 1.987 \times 10^{33} {\rm g}, \nonumber \\
 m_{\rm B} & = & 1.66 \times 10^{-24} {\rm g}. \nonumber
\end{eqnarray}

Note that some of the above constants differ slightly from the ones
used in the papers where the three codes were first presented.

\subsubsection{Interpolation of Tabulated Equation-of-State Data }

For realistic equations of state, the energy density,  pressure
and other thermodynamical quantities are given in tables. 
Intermediate values need to be obtained by a method of interpolation.
We will use two different interpolation schemes, the four-point Lagrange 
interpolation (hereafter LI) and the cubic Hermite interpolation (HI)  (Swesty 1996): 

A) Lagrange interpolation

Let us assume that there is a table which relates the variable $x$
to the variable $y$ at $n$ points, i.e. a set of values $(x_i, y_i)$
for $ i = 1, \dots, n$ are tabulated.
For the LI scheme the interpolated formula $y_{\rm LI}$ can be expressed as
\begin{equation}
y_{\rm LI}(x) \equiv \sum_{i = 1}^n y_i \frac{p_i(x)} {p^{'}(x)|_{x=x_i}},
\end{equation}
where
\begin{eqnarray}
p(x)   & \equiv & (x-x_1) (x-x_2) \dots (x-x_n),  \\
p_i(x) & \equiv & \frac{p(x)} {(x-x_i)}.
\end{eqnarray}
The prime denotes the differentiation with respect to the argument.
This scheme does not guarantee the values of derivatives at the points
in the table. 
In this paper we use the four point Lagrange interpolation, i.e. $n = 4$.

B) Hermite interpolation

In the Hermite interpolation, the interpolated formula $y_{\rm HI}$
for $x_i \le x \le x_{i+1}$ is expressed as
\begin{eqnarray}
y_{\rm HI}(x) & \equiv & y_i h_0(w) + y_{i+1} h_0(1-w) \nonumber \\ &&
 + \left( \frac{dy}{dx} \right)_i (x_{i+1} - x_i) h_1(w) \nonumber \\ &&
 - \left( \frac{dy}{dx} \right)_{i+1} (x_{i+1} - x_i) h_1(1-w),
\end{eqnarray}
where
\begin{equation}
 w     \equiv \frac {x - x_i} {x_{i+1} - x_i}.
\end{equation}
Here $h_0$ and $h_1$ are the cubic Hermite functions defined by
\begin{eqnarray}
h_0(w)   & \equiv & 2 w^3 - 3 w^2 + 1,   \\
h_1(w)   & \equiv & w^3 - 2 w^2 + w .
\end{eqnarray}
It is noted that these cubic Hermite functions are unique polynomials
of degree three satisfying the following relations:
\begin{eqnarray}
        & & h_0(0) = 1, \\
        & & h_0(1) =  h'_0(0) = h'_0(1) = 0, \\
        & & h'_1(0) = 1, \\
        & & h_1(0) = h_1(1) = h'_1(1) = 0. 
\end{eqnarray}
Contrary to the LI, in the Hermite interpolation the values as well as 
their first derivatives at mesh points are exactly reproduced by 
the interpolation formula. 
The main advantage of the Hermite interpolation is that it
preserves the thermodynamical consistency (Swesty 1996).

\section{Equations of state}

\subsection{Relativistic polytropes}

We use the following relation as a polytropic equation of state 
(Tooper \cite{toop65}):
\begin{eqnarray} 
  \varepsilon & = & K {\rho^\gamma \over \gamma - 1} + \rho \, c^2 , \\
            p & = & K\,  \rho^\gamma \ , \\
       \gamma & = &  1 + {1 \over N} \ ,  
\end{eqnarray}
where $K$ and $N$ are the polytropic constant and polytropic index, respectively, while
$\rho$ is the rest mass density.  

It should be noted that this equation of state includes the limiting
case of $\varepsilon = \rho c^2 = constant$, when $\gamma = \infty$ ($N=0$).
The constant density models are also called homogeneous models.
For polytropes of index $N<1.0$, the density does not go to zero smoothly
at the surface and the first derivatives of the density across the surface are discontinuous.
This kind of discontinuity may become the cause of unfavorable behavior
of solutions, unless it is treated carefully.  For constant density models, the
situation is even worse, since the density itself is discontinuous across the surface.

Although polytropic EOSs are not as realistic as tabulated EOSs (but one can 
reproduce neutron star bulk properties with an $N \simeq 1.0 $ polytrope), 
they are helpful  to check numerical codes. Since the
hydrostationary equation can be analytically integrated and no
additional numerical errors arise in solving it.

\subsection{Short description of realistic equations of state}

As discussed in the introduction,  the main uncertainties about 
neutron star properties are related to the unknown interactions of the neutron star 
matter at high density regions. In the last decades, many equations of state 
have been proposed by considering different kinds of interactions into account. 
A large collection of representative equations of state were compiled by Arnett \& 
Bowers~(1977), who constructed nonrotating neutron star models and 
obtained physical quantities for slowly rotating neutron stars. 
We will choose three equations of state of Arnett \& Bowers' compilation,
i.e. equations C, G and L according to their notation. Equations C, G 
and L are those derived by Bethe \& Johnson~(1974), Canuto \& Chitre~(1974) 
and Pandharipande \& Smith~(1975) (see also Pandharipande et al. 
\cite{pand76}), respectively. 
Those equations of state were also used by Friedman et al.~(1986) for 
constructing rapidly rotating relativistic neutron stars models. 

In addition to these equations of state we also employ the WFF3
(UV$_{14}$+TNI) equation of state by Wiringa et al.~(1988), the FPS
equation of state by Lorenz et al.~(1993), and the equation of state
which represents a causal limit (CLES).

Some characteristic features of each equation of state can be summarized 
as follows.

\paragraph{Bethe -- Johnson~I (C):} EOS C is of intermediate stiffness.
The maximum gravitational mass of a spherical neutron star for this
EOS is $1.85M_{\odot}$.  The density range is from $1.71 \times
10^{14}$ g cm$^{-3}$ to $3.23 \times 10^{15}$ g cm$^{-3}$. Hyperons as
well as nucleons are taken into account. The interaction is assumed
non-relativistic and represented by the modified Reid soft core
potential with non-integer parameters. To include the many-body
theory, the constrained variational principle is employed. This
equation of state is joined to the composite BBP($\varepsilon/c^2 >
4.3 \times 10^{11}$g cm$^{-3}$) -- BPS($10^4$ g cm$^{-3} <
\varepsilon/c^2 < 4.3 \times 10^{11}$g cm$^{-3}$) -- FMT($ \varepsilon
/c^2 < 10^4$ g cm$^{-3}$ ). Here BBP, BPS and FMT denote equations of
state by Baym et al.~(1971a), Baym et al.~(1971b) and Feynman et
al.~(1949), respectively.

\paragraph{Canuto -- Chitre (G):} EOS G an extremely soft equation of state.
The maximum gravitational mass of a spherical neutron star for this
equation of state is $1.36M_{\odot}$, so this EOS is not acceptable as
a realistic candidate for the true EOS of neutron star matter. It is
used in this comparison, because it is close to the softest possible
realistic EOS consistent with observational constraints.  The density
range is from $2.37 \times 10^{15}$ g cm$^{-3}$ to $7.23 \times
10^{15}$ g cm$^{-3}$. Crystallization of neutrons is included. The
interaction is non-relativistic and represented by the modified Reid
soft core potential. This equation of state is joined to the composite
PC($7 \times 10^{14}$ g cm$^{-3} < \varepsilon/c^2 < 2.4 \times
10^{15}$g cm$^{-3}$) -- BBP($4.3 \times 10^{11}$ g cm$^{-3} <
\varepsilon/c^2 < 7 \times 10^{14}$g cm$^{-3}$) -- BPS($10^4$ g
cm$^{-3} < \varepsilon/c^2 < 4.3 \times 10^{11}$g cm$^{-3}$) --
FMT($\varepsilon/c^2 < 10^4$ g cm$^{-3}$ ).  Here PC denotes the
equation of state by Pandharipande~(1971).

\paragraph{Pandharipande -- Smith (L):} EOS L is an extremely stiff 
EOS. The maximum gravitational mass of a spherical neutron star for
this equation of state is $2.70M_{\odot}$.  The density range is
larger than $4.386 \times 10^{11}$ g cm$^{-3}$.  Compositions consist
of neutrons. The interaction is assumed non-relativistic and is
represented by the nuclear attraction due to scalar particle exchange.
This equation of state is joined to the BPS ($10^4$ g cm$^{-3} <
\varepsilon/c^2 < 4.4 \times 10^{11}$g cm$^{-3}$) --
FMT($\varepsilon/c^2 < 10^4$ g cm$^{-3}$ ).

\paragraph{Wiringa -- Fiks -- Farbrocini (WFF3):} EOS WFF3 (Wiringa et al., 1988)
is of intermediate stiffness.  At present, the WFF3 equation of state
is regarded as one of the best candidates for the high density region.
This EOS is an improved version of the equation of state by Friedman
\& Pandharipande~(1981).  The nucleon-nucleon interaction described by
a two-body Urbana UV$_{14}$ potential and the phenomenological
three-nucleon TNI interaction are taken into account. Compositions are
considered to be neutrons.  The maximum gravitational mass of a
spherical neutron star for this equation of state is $1.84M_{\odot}$.
Although the usual WFF3 EOS is joined to EOS NV (Negele \& Vautherin,
1973), we will also use a different version, in which it is joined to
the more modern FPS EOS (Lorenz et al., 1993).

\paragraph{Lorenz -- Ravenhall -- Pethick (FPS):} This EOS
is also a modern version of the equation of state by Friedman \&
Pandharipande~(1981). The nucleon-nucleon interaction described by a
two-body Urbana UV$_{14}$ potential and the phenomenological
three-nucleon TNI interaction are taken into account. In the FPS
equation of state the Skyrme model is used, where the effective
interaction has the spatial character of a two-body delta function
plus derivatives.  The FPS equation of state can be considered to be
an improved version of the BBP equation of state in the region of the
lower density.

\paragraph{Causal limit equation of state (CLES):} As an extreme case,
we consider an equation of state which consists of a causal limit EOS
($\varepsilon = p + constant$) for $\varepsilon/c^2 > 1.66 \times
10^{14}$ g cm$^{-3}$ and the FPS EOS below that density. The causal
limit EOS has the property that, in the interior of the star, the phase
velocity of sound is equal to the velocity of light in vacuo, i.e.
$v_s = \sqrt{dp/d \varepsilon}=c$.

\begin{table*}[t]
\caption{\label{t:poly1} Polytropic models}
\begin{tabular}{c c c c c c c c c c}
\hline
Model                      &  N05sn    & N05rn    & N05sr    &  N05mr    &  N05rr  & N075sn    & N075rn    & N075sr  &  N075mr       \\
$ N $                      &  0.5      & 0.5      & 0.5      &  0.5      &  0.5    & 0.75      & 0.75      & 0.75    &  0.75     \\
\hline
$\bar \varepsilon_{\rm c}$ &  1.00e-03 & 1.00e-03 & 1.20e+00 &  1.04e+00 &  2.00e+00 &  3.00e-05 &  3.00e-05 &  8.00e-01 &  6.50e-01 \\
$r_p/r_e$                  &  9.80e-01 & 5.99e-01 & 9.85e-01 &  5.50e-01 &  7.57e-01 &  9.80e-01 &  6.82e-01 &  9.79e-01 &  5.66e-01 \\
$\bar \Omega$              &  6.91e-03 & 2.85e-02 & 2.14e-01 &  9.78e-01 &  1.00e+00 &  1.10e-03 &  4.11e-03 &  1.77e-01 &  6.30e-01 \\
                           &  6.92e-03 &          & 2.19e-01 &  9.80e-01 &           &           &           &  1.79e-01 &  6.31e-01 \\
$\bar M$                   &  6.30e-08 & 9.43e-08 & 1.26e-01 &  1.53e-01 &  1.35e-01 &  3.73e-07 &  4.84e-07 &  1.40e-01 &  1.64e-01 \\
                           &           &          & 1.27e-01 &  1.54e-01 &  1.36e-01 &           &           &  1.41e-01 &  1.65e-01 \\
$\bar M_0$                 &  6.30e-08 & 9.43e-08 & 1.52e-01 &  1.83e-01 &  1.59e-01 &  3.73e-07 &  4.84e-07 &  1.60e-01 &  1.87e-01 \\
                           &           &          & 1.53e-01 &           &  1.60e-01 &           &           &           &           \\
$\bar R_{circ}$            &  3.04e-02 & 4.20e-02 & 4.03e-01 &  5.35e-01 &  4.03e-01 &  1.95e-01 &  2.47e-01 &  5.33e-01 &  7.37e-01 \\
                           &           &          &          &           &           &           &           &           &  7.38e-01 \\ 
$\bar J$                   &  1.31e-13 & 1.47e-12 & 2.16e-03 &  1.72e-02 &  1.07e-02 &  4.55e-12 &  3.35e-11 &  2.87e-03 &  1.75e-02 \\
                           &  1.32e-13 &          & 2.21e-03 &           &           &  4.57e-12 &           &  2.89e-03 &           \\
$\bar I$                   &  1.90e-11 & 5.17e-11 & 1.01e-02 &  1.76e-02 &  1.07e-02 &  4.16e-09 &  8.15e-09 &  1.62e-02 &  2.77e-02 \\
$T/W$                      &  5.18e-03 & 1.26e-01 & 5.01e-03 &  1.47e-01 &  8.76e-02 &  4.93e-03 &  8.96e-02 &  5.94e-03 &  1.11e-01 \\
                           &  5.20e-03 &          & 5.15e-03 &  1.49e-01 &  8.96e-02 &  4.96e-03 &           &  5.95e-03 &  1.12e-01 \\
$Z_p$                      &  2.10e-06 & 3.10e-06 & 6.32e-01 &  9.64e-01 &  9.97e-01 &  1.94e-06 &  2.55e-06 &  4.59e-01 &  6.10e-01 \\
                           &           &          & 6.42e-01 &  9.75e-01 &  10.0e-01 &           &  2.56e-06 &  4.61e-01 &  6.13e-01 \\
$Z^f_{eq}$                 & -2.08e-04 &-1.19e-03 & 4.10e-01 & -3.62e-01 & -1.42e-01 & -2.12e-04 & -1.01e-03 &  2.61e-01 & -3.19e-01 \\
                           & -2.09e-04 &          & 4.15e-01 & -3.64e-01 & -1.47e-01 & -2.13e-04 &           &  2.65e-01 & -3.20e-01 \\
$Z^b_{eq}$                 &  2.12e-04 & 1.20e-03 & 8.66e-01 &  2.99e+00 &  2.86e+00 &  2.16e-04 &  1.02e-03 &  6.59e-01 &  1.72e+00 \\
                           &  2.13e-04 &          & 8.90e-01 &  3.02e+00 &  2.92e+00 &  2.17e-04 &           &  6.70e-01 &  1.73e+00 \\
$e$                        &  1.99e-01 & 8.01e-01 & 1.31e-01 &  6.95e-01 &  5.64e-01 &  1.96e-01 &  7.32e-01 &  1.66e-01 &  7.02e-01 \\
                           &  2.05e-01 &          & 1.72e-01 &  7.02e-01 &  5.74e-01 &  2.09e-01 &           &  1.97e-01 &  7.06e-01 \\
\hline   
\end{tabular}
\end{table*}

%%%%%%%%%%%%%%%%%%%%%%%%%%%%%%%%%%%%%%%%%%%%%%%%%%%%%%%%%%%%%%%%%%%%%%%%%%%%%%
\begin{table*}[t]
\caption{\label{t:poly2} Polytropic models (continued)} 
\begin{tabular}{c c c c c c c c c c c}
\hline
Model                      &  N10sn    &  N10rn    &  N10sr    &  N10mr    &  N10rr    &  N15sn    &  N15rn    &  N15sr    &  N15mr    &  N15rr    \\
$ N $                      &  1.0      &  1.0      &  1.0      &  1.0      &  1.0      &  1.5      &  1.5      &  1.5      &  1.5      &  1.5      \\
\hline
$\bar \varepsilon_{\rm c}$ &  1.00e-06 &  1.00e-06 &  4.00e-01 &  3.40e-01 &  1.00e+00 &  1.00e-09 &  1.00e-09 &  6.50e-01 &  6.10e-02 &  1.50e-01 \\
$r_p/r_e$                  &  9.76e-01 &  6.39e-01 &  9.72e-01 &  5.84e-01 &  8.34e-01 &  9.60e-01 &  7.08e-01 &  8.68e-01 &  6.20e-01 &  8.40e-01 \\
$\bar \Omega$              &  2.00e-04 &  7.00e-04 &  1.26e-01 &  3.77e-01 &  4.00e-01 &  6.53e-06 &  1.58e-05 &  1.61e-01 &  1.11e-01 &  1.20e-01\\
                           &           &           &           &           &  4.01e-01 &           &           &           &           &           \\
$\bar M$                   &  2.54e-06 &  3.29e-06 &  1.65e-01 &  1.88e-01 &  1.61e-01 &  9.76e-05 &  1.11e-04 &  2.10e-01 &  2.91e-01 &  2.66e-01 \\
                           &           &           &  1.66e-01 &  1.89e-01 &           &           &           &           &           &           \\
$\bar M_0$                 &  2.54e-06 &  3.29e-06 &  1.82e-01 &  2.07e-01 &  1.73e-01 &  9.76e-05 &  1.11e-04 &  2.08e-01 &  3.04e-01 &  2.77e-01 \\
$\bar R_{circ}$            &  1.27e+00 &  1.71e+00 &  7.82e-01 &  1.09e+00 &  6.79e-01 &  5.29e+01 &  6.63e+01 &  1.27e+00 &  2.85e+00 &  1.79e+00 \\
$\bar J$                   &  2.14e-10 &  1.50e-09 &  4.24e-03 &  2.02e-02 &  9.48e-03 &  3.60e-07 &  1.26e-06 &  1.02e-02 &  3.88e-02 &  2.13e-02 \\
                           &           &           &  4.25e-03 &           &           &           &           &           &           &           \\
$\bar I$                   &  1.07e-06 &  2.14e-06 &  3.36e-02 &  5.36e-02 &  2.37e-02 &  5.51e-02 &  7.99e-02 &  6.32e-02 &  3.50e-01 &  1.78e-01 \\
                           &           &           &           &  5.37e-02 &           &           &  8.00e-02 &  6.33e-02 &           &           \\
$T/W$                      &  5.56e-03 &  9.10e-02 &  6.61e-03 &  8.36e-02 &  3.48e-02 &  7.48e-03 &  5.24e-02 &  1.32e-02 &  4.75e-02 &  2.29e-02 \\
                           &  5.57e-03 &           &  6.63e-03 &  8.41e-02 &  3.52e-02 &           &           &           &  4.76e-02 &  2.30e-02 \\
$Z_p$                      &  2.04e-06 &  2.71e-06 &  3.23e-01 &  4.04e-01 &  4.56e-01 &  1.91e-06 &  2.25e-06 &  2.57e-01 &  1.94e-01 &  2.29e-01 \\
                           &  2.05e-06 &  2.74e-06 &  3.26e-01 &  4.05e-01 &  4.58e-01 &           &  2.27e-06 &  2.58e-01 &  1.95e-01 &  2.30e-01 \\
$Z^f_{eq}$                 & -2.52e-04 & -1.19e-03 &  1.54e-01 & -2.83e-01 & -5.95e-02 & -3.44e-04 & -1.05e-03 & -5.33e-02 & -2.24e-01 & -8.10e-02 \\
                           &           &           &  1.56e-01 & -2.84e-01 & -6.19e-02 &           &           & -5.43e-02 & -2.25e-01 & -8.17e-02 \\
$Z^b_{eq}$                 &  2.56e-04 &  1.20e-03 &  4.97e-01 &  1.15e+00 &  1.03e+00 &  3.48e-04 &  1.05e-03 &  5.75e-01 &  6.23e-01 &  5.49e-01 \\
                           &  2.57e-04 &           &  5.02e-01 &           &  1.04e+00 &           &           &  5.78e-01 &           &  5.51e-01 \\
$e$                        &  2.16e-01 &  7.70e-01 &  2.03e-01 &  7.14e-01 &  4.75e-01 &  2.80e-01 &  7.06e-01 &  4.44e-01 &  7.30e-01 &  4.93e-01 \\
                           &  2.27e-01 &           &  2.25e-01 &  7.18e-01 &  4.85e-01 &  2.84e-01 &  7.07e-01 &  4.52e-01 &           &  5.00e-01 \\
\hline   
\end{tabular}
\end{table*}

\section{Results}

\subsection{Models selected for polytropes}

We have started our comparison project by selecting several
representative polytropic models, the parameters of which are shown in
Tables~\ref{t:poly1}-\ref{t:poly2}.  We have chosen
\begin{enumerate}
\item models of very low central density (nearly Newtonian) with slow 
 and rapid rotation, 
\item models of high central density (relativistic) with slow and
  rapid rotation, and
\item models at the maximum mass for each EOS.
\end{enumerate}

Note that, the maximum mass model almost coincides with the maximum
angular velocity model, unless there is a large phase transition at
densities close to the central density of the maximum mass star (cf.
Cook, Shapiro \& Teukolsky, 1994b and Stergioulas \& Friedman, 1995).
For all EOSs in this comparison the two models almost coincide.

In order to evaluate the performance of our numerical codes for models
with discontinuous density distribution, we also compare a number of
homogeneous models which cover both highly relativistic and Newtonian, rapidly
rotating and nonrotating cases,  as shown in Tables~\ref{t:spher_const} and 
\ref{t:rot_const}
(the contents of these tables will be described in Sect.~\ref{s:tables}).

%%%%%%%%%%%%%%%%%%%%%%%%%%%%%%%%%%%%%%%%%%%%%%%%%%%%%%%%%%%%%%%%%%%%%%%%%%%%%%

\subsection{Models selected for realistic equations of state}

As discussed in the previous section, we use six representative
realistic EOSs: C, G, L, WFF3+FPS, WFF3+NV and FPS. In addition, we use
the causal limit EOS CLES. For each equation of state, we compute
several models as shown Tables~\ref{t:spher_real} -- \ref{t:rot_real2}. 
The models correspond to
the maximum mass model, a fast rotating $1.4 M_\odot$ model and a nonrotating
model for each EOS.

\subsection{Computed Quantities}

\subsubsection{Grid and physical quantities}

For this comparison project, KEH(OR) and KEH(SF) have used grids with
($71 \times 201) $ and ($261 \times 401$) (angular$\times$radial)
grid-points. In the equatorial plane, half of the radial grid-points
are inside the star. BGSM uses 21 $\times$ 41 or 33 $\times$ 65 grid
points (note that the notion of ``grid points'' is not very significant
for a spectral method; the above numbers should better be referred to as
the numbers of basis functions employed in the expansions of the
physical fields).

Here we summarize the notation of computed physical quantities:
\begin{flushleft}
\begin{center}
\begin{tabular}{l p{12truecm}}
$\varepsilon_c$     & Central energy density \\
$r_p/r_{e} $          & Ratio of polar to equatorial radii \\
$\Omega$            & Angular velocity of the star \\
%%  $\Omega_{\rm p}$ & Angular velocity of a particle on its maximum 
%%  circular orbit \\
%$\Omega_{\rm K}$    & Kepler angular velocity \\
%%$P$                 & Rotational period \\
$M_0$               & Baryon mass \\
$M_p$               & Proper mass \\
$M$                 & Gravitational mass \\
$R_{circ}$            & Equatorial circumferential radius \\
$r_{e}$               & Equatorial coordinate radius \\
$J$                 & Total angular momentum \\ 
$I$                 & Moment of inertia about the rotation axis \\
$T$                 & Rotational energy \\
$W$                 & Gravitational energy \\
%% $e^{\nu_c}$         & Central value of the metric component $e^{\nu}$ \\
%% $\omega_c$          & Central value of the dragging of the inertial frame \\
$v_e$               & Velocity of comoving observer at the equator \\ 
$ $                    & relative to the locally non-rotating observer  \\ 
$Z_p$               & Polar redshift \\
$Z_c$               & Central redshift \\
$Z_{eq}^b$             & Equatorial redshift in the backward direction \\
$Z_{eq}^f$             & Equatorial redshift in the forward direction \\
$e$                 & Intrinsic eccentricity of the star's surface \\
$GRV2$              & Two dimensional virial identity \\
$GRV3$              & Three dimensional virial identity\\
\end{tabular} 
\end{center}
\end{flushleft}

Some of the quantities in the above list can be expressed as follows:
\begin{eqnarray}
Z_p   & = & e^{-2 \nu_p} - 1, \\
Z_{eq}^f & = & \left({1 - v_e \over 1 + v_e} \right)^{1/2} {e^{-\nu_e} 
\over 1 + r_e e^{(\nu_e-\beta_e)/2} \omega_e} - 1, \\
Z_{eq}^b & = & \left({1 + v_e \over 1 - v_e} \right)^{1/2} {e^{-\nu_e} 
\over 1 - r_e e^{(\nu_e-\beta_e)/2} \omega_e} - 1, 
\end{eqnarray}
where subscripts $_p$ and $_e$ denote values at the pole
and the equatorial surface, respectively.
\begin{equation}
M_0 = 2 \pi \int \rho {e^{2 \alpha + \beta} \over \sqrt{ 1- v^2}}
r^2 \sin \theta dr d\theta,
\end{equation}  
\begin{equation}
M_p = 2 \pi \int \varepsilon {e^{2 \alpha + \beta} \over \sqrt{ 1- v^2}}
r^2 \sin \theta dr d\theta,
\end{equation}  
\begin{eqnarray}
M &=& 2 \pi \int \Biggl[ e^{2 \alpha + \beta} 
\left\{ {(\varepsilon + p)(1+v^2) \over 1-v^2} + 2 p \right\} \nonumber \\
 && + 2 r \sin \theta \omega e^{\beta} {(\varepsilon + p) v \over 1-v^2} \Biggr] 
r^2 \sin \theta dr d\theta,
\end{eqnarray}
\begin{equation}
J = 2 \pi \int e^{2\alpha + 2 \beta} {(\varepsilon + p) v \over 1-v^2}
r^3 \sin^2 \theta dr d\theta,
\end{equation}  
\begin{equation}
T = {1 \over 2} \int \Omega dJ = 2 \pi \int e^{2\alpha + 2 \beta} 
{(\varepsilon + p) v \over 1-v^2} \Omega r^3 \sin^2 \theta dr d\theta,
\end{equation}  
\begin{equation}
W = M_p c^2 +  T - M c^2,
\end{equation}  
and
\begin{equation}
I = { J \over \Omega}.
\end{equation}

The eccentricity of the meridional cross section is defined by the following
procedure (Friedman et al. \cite{frie86}). If the surface of the star is defined by
\begin{equation}
 r = r_s(\theta), 
\end{equation}
the metric of the stellar surface can be expressed as 
\begin{equation}
d \sigma^2 = e ^{2 \beta} r^2 \sin^2 \theta d \varphi^2
+ e^{2 \alpha} \left[ \left( {d r_s \over d \theta} \right)^2
+ r_s(\theta)^2 \right] d \theta^2.
\end{equation}
If we embed this surface in the flat three dimensional space, 
it is expressed as
\begin{equation}
R = R_s(z),
\end{equation}
in cylindrical coordinates $(R, \varphi, z)$.
The 2-metric of this surface is
\begin{equation}
 d \sigma^2 = \left[ \left({ d R_s \over dz} \right)^2 + 1 \right] dz^2
  + R_s^2 d \varphi^2.
\end{equation}
Comparing these two equations, we have the following relations,
if they express the same surface geometry:
\begin{equation}
   R_s(\theta) = e^{\beta} r \sin \theta,
\end{equation}
and
\begin{eqnarray}
   z_s(\theta) = \int_\theta^{\pi/2} d \theta & \Biggl\{ & e^{2 \alpha} \left[ 
  \left( {d r_s \over d \theta} \right)^2 + r_s(\theta)^2 \right] \nonumber \\ 
  &&  - \left({d R_s \over d \theta} \right)^2 \Biggr\}^{1/2}.
\end{eqnarray}
Using these quantities the eccentricity is defined as
\begin{equation}
e \equiv \sqrt{ 1 - \left({ z_s(\theta = 0) \over R_s(\theta=\pi/2)}
\right)^2 }.
\label{eccen}
\end{equation}

For polytropes, it is convenient to express quantities in dimensionless 
form, by using $K^{N/2}$ as a fundamental length scale, as was done in 
Cook et al.~(1994a). In geometrized units ($c = G = 1$),  
dimensionless quantities are define as follows:
%
%\begin{mathletters}
\begin{eqnarray}
\bar r           & \equiv & K^{-N/2} r, \\
\bar R_{circ}    & \equiv & K^{-N/2} R_{circ}, \\
\bar \Omega      & \equiv & K^{N/2} \Omega, \\
\bar \varepsilon & \equiv & K^N \varepsilon, \\
\bar p           & \equiv & K^N p, \\
\bar \rho        & \equiv & K^N \rho, \\
\bar J           & \equiv & K^{-N} J, \\
\bar I           & \equiv & K^{-3N/2} J, \\
\bar M           & \equiv & K^{-N/2} M, \\
\bar M_0         & \equiv & K^{-N/2} M_0.
\end{eqnarray}
%\end{mathletters}
%

\begin{table}[t]
\caption{\label{t:spher_const} Spherical constant density models}
\begin{tabular}{c c c}
\hline
Model                      &  N00sn    & N00sr     \\
\hline                                           
$\bar \varepsilon_{\rm c}$ &  1.00e+00 &  1.00e+00 \\
$\bar p_{\rm c}$           &  1.00e-04 &  1.00e+00 \\
$r_p/r_e$                  &  1.00e+00 &  1.00e+00 \\
$\bar M$                   &  1.38e-06 &  1.12e-01 \\
                           &  1.41e-06 &  1.13e-01 \\
$\bar M_0$                 &  1.38e-06 &  1.59e-01 \\
                           &  1.41e-06 &  1.61e-01 \\
$\bar R_{circ}$            &  6.91e-03 &  2.99e-01 \\
                           &  6.93e-03 &  3.00e-01 \\
$Z_p$                      &  2.00e-04 &  9.71e-01 \\
                           &  2.04e-04 &  10.1e-01 \\
$Z_{eq}^f$                 &  2.00e-04 &  9.71e-01 \\
                           &  2.04e-04 &  10.1e-01 \\
$Z_{eq}^b$                 &  2.00e-04 &  9.71e-01 \\
                           &  2.04e-04 &  10.1e-01 \\
\hline
\end{tabular}
\end{table}
%%%%%%%%%%%%%%%%%%%%%%%%%%%%%%%%%%%%%%%%%%%%%%%%%%%%%%%%%%%%%%%%%%%%%%%%%%%%%%
\begin{table}
\caption{\label{t:rot_const} Rotating constant density models}
\begin{tabular}{c c c}
\hline
Model                      &  N00rn    &  N00rr    \\
\hline                                           
$\bar \varepsilon_{\rm c}$ &  1.00e+00 &  1.00e+00 \\
$\bar p_{\rm c}$           &  1.00e-04 &  1.00e+00 \\
$r_p/r_e$                  &  6.50e-01 &  7.00e-01 \\
$\bar \Omega$              &  1.00e+00 &  1.40e+00 \\
                           &  1.02e+00 &  1.41e+00 \\
$\bar M$                   &  2.05e-06 &  1.35e-01 \\
                           &  2.06e-06 &  1.39e-01 \\
$\bar M_0$                 &  2.05e-06 &  1.84e-01 \\
                           &  2.06e-06 &  1.87e-01 \\
$\bar R_{circ}$            &  9.06e-03 &  3.45e-01 \\
                           &  9.11e-03 &  3.46e-01 \\
$\bar J$                   &  6.76e-11 &  1.37e-02 \\
                           &  6.97e-11 &  1.41e-02 \\
$\bar I$                   &  6.74e-11 &  9.82e-03 \\
                           &  6.87e-11 &  10.0e-03 \\
$T/W$                      &  1.08e-01 &  1.63e-01 \\
                           &  1.14e-01 &  1.68e-01 \\
$Z_p$                      &  2.83e-04 &  1.60e+00 \\
                           &  2.86e-04 &  1.71e+00 \\
$Z_{eq}^f$                 & -8.81e-03 & -1.55e-01 \\
                           & -8.96e-03 & -1.60e-01 \\
$Z_{eq}^b$                 &  9.38e-03 &  9.41e+00 \\
                           &  9.54e-03 &  11.3e+00 \\
$e$                        &  7.60e-01 &  7.07e-01 \\
                           &  7.61e-01 &  7.11e-01 \\
\hline
\end{tabular}
\end{table}

\begin{table*}
\caption{\label{t:spher_real} Spherical models with realistic EOSs}
\begin{tabular}{c c c c c c c c}
\hline
Model                              &  Csr     &  Gsr     &   Lsr    &WFF(FPS)sr &WFF(NV)sr &  FPSsr     &  CLESsr   \\
EOS                                &  C       &  G       &   L      & WFF3+FPS  & WFF3+NV  &   FPS     &  CLES     \\
\hline
$\varepsilon_{\rm c}$[g cm$^{-3}]$ & 1.09e+15 & 6.31e+15 & 4.30e+14 & 1.22e+15 & 1.22e+15 &  1.31e+15 &  1.85e+14\\
$r_p/r_e$                          & 1.00e+00 & 1.00e+00 & 1.00e+00 & 1.00e+00 & 1.00e+00 &  1.00e+00 &  1.00e+00 \\
$M[M_{\odot}]$                     & 1.41E+00 & 1.36E+00 & 1.39e+00 & 1.41e+00 & 1.41e+00 &  1.41e+00 &  1.41e+00 \\
                                   &          & 1.37E+00 & 1.41e+00 &          &          &           &           \\
$M_0[M_{\odot}]$                   & 1.55e+00 & 1.57e+00 & 1.49e+00 & 1.57e+00 & 1.57e+00 &  1.57e+00 &  1.51e+00 \\
                                   &          &          & 1.53e+00 &          &          &           &           \\
$R_{circ}$[km]                     & 1.19e+01 & 6.94e+00 & 1.48e+01 & 1.09e+01 & 1.09e+01 &  1.08e+01 &  1.77e+01 \\
$Z_p$                              & 2.39e-01 & 5.28e-01 & 1.76e-01 & 2.69e-01 & 2.68e-01 &  2.72e-01 &  1.43e-01 \\
                                   & 2.41e-01 & 5.38e-01 & 1.80e-01 & 2.71e-01 & 2.71e-01 &  2.74e-01 &  1.44e-01 \\
$Z_{eq}^f$                         & 2.39e-01 & 5.28e-01 & 1.76e-01 & 2.69e-01 & 2.68e-01 &  2.72e-01 &  1.43e-01 \\
                                   & 2.41e-01 & 5.38e-01 & 1.80e-01 & 2.71e-01 & 2.71e-01 &  2.74e-01 &  1.44e-01 \\
$Z_{eq}^b$                         & 2.39e-01 & 5.28e-01 & 1.76e-01 & 2.69e-01 & 2.68e-01 &  2.72e-01 &  1.43e-01 \\
                                   & 2.41e-01 & 5.38e-01 & 1.80e-01 & 2.71e-01 & 2.71e-01 &  2.74e-01 &  1.44e-01 \\
\hline                                                           
\end{tabular}
\end{table*}

\begin{table*}
\caption{\label{t:rot_real1} Rotating models with realistic EOSs}
\begin{tabular}{c c c c c c c c c}
\hline
Model                              &    Cbr    &    Cmr    &   Gbr     &    Gmr    &    Lbr    &  L(L)mr   &  L(H)mr   & WFF(FPS)br \\
EOS                                &    C      &    C      &   G       &    G      &    L      &   L(LI)   &  L(HI)    & WFF3+FPS   \\
\hline
$\varepsilon_{\rm c}$[g cm$^{-3}$] &  8.70e+14 &  2.64e+15 &  3.10e+15 &  5.58e+15 &  3.90e+14 &  1.20e+15 &  1.20e+15 &  9.70e+14 \\
$r_p/r_e$                          &  6.73e-01 &  5.72e-01 &  6.37e-01 &  5.77e-01 &  7.08e-01 &  5.53e-01 &  5.53e-01 &  6.30e-01 \\
$\Omega$[$10^4$ s$^{-1}$]          &  5.89e-01 &  1.07e+00 &  1.16e+00 &  1.57e+00 &  4.21e-01 &  8.17e-01 &  8.13e-01 &  7.00e-01 \\
                                   &           &           &           &  1.58e+00 &  4.25e-01 &  8.18e-01 &  8.14e-01 &  7.01e-01 \\
$M[M_{\odot}]$                     &  1.41e+00 &  2.17e+00 &  1.41e+00 &  1.57e+00 &  1.41e+00 &  3.31e+00 &  3.29e+00 &  1.42e+00 \\
                                   &           &           &           &           &  1.43e+00 &  3.33e+00 &  3.30e+00 &           \\
$M_0[M_{\odot}]$                   &  1.52e+00 &  2.48e+00 &  1.58e+00 &  1.79e+00 &  1.50e+00 &  3.90e+00 &  3.87e+00 &  1.55e+00 \\
                                   &  1.53e+00 &  2.49e+00 &  1.59e+00 &           &  1.53e+00 &  3.92e+00 &  3.88e+00 &           \\
$R_{circ}$[km]                     &  1.52e+01 &  1.32e+01 &  1.02e+01 &  9.24e+00 &  1.76e+01 &  1.85e+01 &  1.85e+01 &  1.43e+01 \\
                                   &           &  1.34e+01 &           &  9.25e+00 &  1.77e+01 &           &           &           \\
$cJ/GM^2_{\odot}$                  &  1.20e+00 &  2.99e+00 &  1.19e+00 &  1.54e+00 &  1.24e+00 &  7.72e+00 &  7.62e+00 &  1.31e+00 \\
                                   &           &  3.00e+00 &           &           &  1.28e+00 &  7.80e+00 &  7.64e+00 &           \\
$I$[$10^{45}$ g cm$^2$]            &  1.80e+00 &  2.45e+00 &  9.01e-01 &  8.56e-01 &  2.60e+00 &  8.31e+00 &  8.23e+00 &  1.64e+00 \\
                                   &           &  2.46e+00 &  9.03e-01 &  8.59e-01 &  2.64e+00 &  8.40e+00 &  8.25e+00 &  1.65e+00 \\
$T/W$                              &  8.44e-02 &  1.10e-01 &  9.59e-02 &  1.05e-01 &  8.38e-02 &  1.37e-01 &  1.36e-01 &  1.01e-01 \\
                                   &  8.49e-02 &           &  9.69e-02 &  1.07e-01 &  8.47e-02 &  1.39e-01 &           &  1.02e-01 \\
$Z_p$                              &  2.46e-01 &  6.86e-01 &  4.63e-01 &  7.44e-01 &  1.90e-01 &  8.26e-01 &  8.18e-01 &  2.82e-01 \\
                                   &           &  6.89e-01 &  4.66e-01 &  7.49e-01 &  1.93e-01 &  8.31e-01 &  8.21e-01 &  2.84e-01 \\
$Z_{eq}^f$                         & -1.81e-01 & -3.30e-01 & -2.42e-01 & -3.36e-01 & -1.42e-01 & -3.47e-01 & -3.46e-01 & -2.10e-01 \\
                                   &           & -3.31e-01 & -2.43e-01 & -3.38e-01 & -1.43e-01 & -3.49e-01 & -3.47e-01 & -2.11e-01 \\
$Z_{eq}^b$                         &  6.96e-01 &  1.94e+00 &  1.28e+00 &  2.10e+00 &  5.44e-01 &  2.43e+00 &  2.41e+00 &  8.11e-01 \\
                                   &  6.98e-01 &  1.96e+00 &           &  2.12e+00 &           &  2.46e+00 &  2.42e+00 &  8.12e-01 \\
$e$                                &  6.84e-01 &  6.90e-01 &  6.75e-01 &  6.74e-01 &  6.65e-01 &  6.97e-01 &  7.02e-01 &  7.13e-01 \\
                                   &  6.88e-01 &  7.26e-01 &  6.82e-01 &  6.81e-01 &  6.70e-01 &  7.04e-01 &  7.03e-01 &  7.18e-01 \\
\hline
\end{tabular}
\end{table*}

\begin{table*}
\caption{\label{t:rot_real2} Rotating models with realistic EOSs (continued)}
\begin{tabular}{c c c c c c c c}
\hline
Model                              & WFF(FPS)mr &  WFF(NV)br & WFF(NV)mr &  FPSbr    &  FPSmr    &  CLESbr   &  CLESmr    \\
                                   &           &           &(max. mass)&           &(max. mass)&           &(max. mass) \\
EOS                                & WFF3+FPS  &  WFF3+NV  &  WFF3+NV  &  FPS      &  FPS      &  CLES     &  CLES      \\
\hline
$\varepsilon_{\rm c}$[g cm$^{-3}$] &  2.70e+15 &  9.70e+14 &  2.71e+15 &  1.02e+15 &  2.91e+15 &  1.81e+14 &  4.20e+14  \\
$r_p/r_e$                          &  5.65e-01 &  6.30e-01 &  5.65e-01 &  6.40e-01 &  5.68e-01 &  5.90e-01 &  5.31e-01  \\
$\Omega$[$10^4$ s$^{-1}$]          &  1.15e+00 &  6.98e-01 &  1.15e+00 &  6.97e-01 &  1.18e+00 &  3.63e-01 &  6.03e-01  \\
                                   &           &  6.99e-01 &           &           &           &           &  6.06e-01  \\
$M[M_{\odot}]$                     &  2.19e+00 &  1.41e+00 &  2.19e+00 &  1.41e+00 &  2.12e+00 &  1.41e+00 &  6.64e+00  \\
                                   &  2.20e+00 &  1.42e+00 &  2.20e+00 &           &  2.13e+00 &  1.42e+00 &  6.71e+00  \\
$M_0[M_{\odot}]$                   &  2.55e+00 &  1.54e+00 &  2.55e+00 &  1.54e+00 &  2.46e+00 &  1.50e+00 &  8.36e+00  \\
                                   &           &  1.55e+00 &           &           &           &           &  8.40e+00  \\
$R_{circ}$[km]                     &  1.28e+01 &  1.43e+01 &  1.28e+01 &  1.41e+01 &  1.24e+01 &  2.30e+01 &  2.86e+01  \\
                                   &           &           &           &           &           &           &  2.87e+01  \\
$cJ/GM^2_{\odot}$                  &  3.22e+00 &  1.31e+00 &  3.21e+00 &  1.27e+00 &  2.96e+00 &  1.62e+00 &  3.61e+01  \\
                                   &  3.23e+00 &           &  3.22e+00 &           &  2.97e+00 &           &  3.64e+01  \\
$I$[$10^{45}$ g cm$^2$]            &  2.45e+00 &  1.64e+00 &  2.44e+00 &  1.60e+00 &  2.20e+00 &  3.92e+00 &  5.25e+01  \\
                                   &  2.46e+00 &  1.65e+00 &  2.46e+00 &           &  2.21e+00 &  3.93e+00 &  5.30e+01  \\
$T/W$                              &  1.22e-01 &  1.00e-01 &  1.22e-01 &  9.70e-02 &  1.17e-01 &  1.13e-01 &  1.92e-01  \\
                                   &  1.23e-01 &  1.01e-01 &  1.23e-01 &  9.76e-02 &  1.18e-01 &           &  1.97e-01  \\
$Z_p$                              &  7.63e-01 &  2.81e-01 &  7.62e-01 &  2.81e-01 &  7.47e-01 &  1.60e-01 &  1.46e+00  \\
                                   &  7.64e-01 &  2.83e-01 &  7.64e-01 &  2.82e-01 &  7.55e-01 &  1.61e-01 &  1.49e+00  \\
$Z_{eq}^f$                         & -3.39e-01 & -2.10e-01 & -3.38e-01 & -2.05e-01 & -3.37e-01 & -1.92e-01 & -4.02e-01  \\
                                   & -3.41e-01 & -2.11e-01 & -3.40e-01 & -2.06e-01 & -3.39e-01 & -1.93e-01 & -4.07e-01  \\
$Z_{eq}^b$                         &  2.19e+00 &  8.10e-01 &  2.18e+00 &  8.03e-01 &  2.15e+00 &  5.24e-01 &  5.89e+00  \\
                                   &  2.21e+00 &  8.12e-01 &  2.20e+00 &  8.05e-01 &  2.17e+00 &  5.25e-01 &  6.14e+00  \\
$e$                                &  6.93e-01 &  7.13e-01 &  6.93e-01 &  7.05e-01 &  6.84e-01 &  7.69e-01 &  7.20e-01  \\
                                   &  7.28e-01 &  7.18e-01 &  7.27e-01 &  7.10e-01 &  6.90e-01 &  7.70e-01 &  7.29e-01  \\
\hline
\end{tabular}
\end{table*}

\subsubsection{Virial Theorem}

Equilibrium configurations in Newtonian gravity satisfy the following relation:
\begin{equation}
 2 T + 3 (\gamma -1) U + W = 0 ,
\end{equation}
where $U$ is the internal energy. This relation is called the virial 
relation and has been used to check the accuracy of numerically 
obtained solutions (see e.g. Ostriker \& Mark \cite{ostr68}, Tassoul \cite{tass78}).

In general relativity, similar relations were first found by Bonazzola~(1973). 
Recently, two virial identities in general relativity have been discovered 
by Gourgoulhon \& Bonazzola~(1994) and Bonazzola \& Gourgoulhon~(1994).
Those identities are valid for a general asymptotically flat spacetime. 
We can use these identities to estimate the numerical error. 
Let us define two quantities $\lambda_2$ and $\lambda_3$ as follows:
%
%%%%%\samepage{         % causes some troubles
\begin{equation}
   \lambda_2 \equiv  \frac{ \displaystyle 8 \pi \int_0^{+\infty}
     \int_0^{\pi} \left[ p + (\varepsilon + p) \frac{v^2}{1 - v^2 }
     \right] e^{2\mu} \, r \, dr\, d\theta } 
  { \displaystyle
     \int_0^{+\infty} \int_0^{\pi} \left[ \partial \nu \partial \nu
     - \frac{3}{4} e^{2\psi-2\nu} \partial \omega \partial \omega 
     \right] \, r \, dr \, d\theta }, 
\end{equation}
and
\begin{eqnarray}
 \lambda_3 &\equiv&  
   4 \pi \int_0^{+\infty} \int_0^{\pi}  \left[ 3 p + (\varepsilon + p)  
      \frac{ v^2}{ 1 - v^2 } \right] e^{2\mu + \psi} \, r  \, dr \, d\theta 
      \nonumber \\ 
 &&  \times \Bigg\{ \int_0^{+\infty} \int_0^{\pi} \bigg[ 
          \partial \nu \partial \nu - \frac{1}{2} \partial \mu \partial \psi  \nonumber \\
 &&        + \frac{e^{2\mu -2\psi}}{2} r \sin^2\theta \Bigl( {\partial \mu 
          \over \partial r}
+ \frac{1}{r \tan\theta} {\partial \mu 
          \over \partial \theta} \Bigr)  \nonumber \\ 
 && + \frac{1}{4 r} \left( 1 - e^{2\mu -2\psi} r^2\sin^2\theta \right)
        \Bigl( {\partial \psi \over \partial r} +\frac{1}{r \tan \theta} 
        {\partial \psi \over \partial \theta} \nonumber \\
&&- \frac{1}{r \sin^2\theta} \Bigr)
        - \frac{3}{8} e^{2\psi-2\nu} \partial \omega \partial \omega \bigg] 
        e^{\psi} \,  r \, dr\, d\theta  \Bigg\} ^{-1},
\end{eqnarray}
%%%%%}
with the abridged notation
\begin{equation}
  \partial \mu \partial \psi \equiv {\partial \mu \over \partial r} 
    {\partial \psi \over \partial r} + {1\over r^2} 
    {\partial \mu \over \partial \theta} 
    {\partial \psi \over \partial \theta}  \ .
\end{equation}
Then, we define: 
\begin{eqnarray}
GRV2 & \equiv  & | 1 - \lambda_2|, \\
GRV3 & \equiv  & | 1 - \lambda_3|.
\end{eqnarray}
If the Einstein equations are satisfied, these quantities satisfy
the following virial identities:
\begin{eqnarray}
   GRV2 & = & 0, \\
   GRV3 & = & 0.
\end{eqnarray}
Since exact solutions for the stationary problems satisfy
the above relations, we can choose GRV2 and GRV3 as the error 
indicators for numerically obtained solutions.
Note that due to its three dimensional character, GRV3 gives a 
larger weight to the external layers of the star. 
GRV3 is a relativistic generalization of the Newtonian virial 
theorem Eq.~(43).

In practice, however, it should be noted that the virial identities 
in the above form are not always close to the accuracy of 
numerical results.  In particular, for GRV2 the integration is 
done by integrating in the $\theta$ coordinate as seen from the 
definition of GRV2. If one does not use the $\theta$ coordinate 
for solving equilibrium structures, one needs to change variables 
and in that procedure accuracy may be lost and the resultant
values may become worse than the ``real" accuracy before the variable change.  

Concerning the quantity GRV3, the metric potentials in the vacuum region 
contribute to the integral considerably. It implies that if
only the finite regions are treated,  as in the KEH(OR)
code, a large portion of the integrand cannot be taken into account,
However, the expressions for the GRV2 and GRV3 are not unique
because we can do the integral by part and replace the second derivatives
with the matter terms, using Einstein's equations. In this way, the contribution 
far away from the star becomes less important. In the KEH(OR) code, GRV2 and GRV3
are evaluated through
\begin{eqnarray}
 \lambda_2' &\equiv&
8 \pi  \int_0^{r_{max}}\int_0^{\pi}
   \left[  p + (\varepsilon+p){v^2 \over (1-v^2)} 
  \right] e^{2\alpha}\, r \, dr\, d\theta \nonumber \\ 
&& \times \Bigg\{
\int^{r_{max}}_0 \int^\pi_0
 {1 \over 2}\Biggl[
  {\nu,_r}^2+{1\over r^2}{\nu,_\theta}^2 \nonumber \\
&& -\nu \left( \nu,_{rr}+{1 \over r}\nu,_r +{1\over r^2}\nu,_{\theta\theta} \right)
 \Biggr]
 r\, dr\, d\theta   \nonumber \\
 & &  - {3\over 4}  \int^{r_{max}}_0 \int^\pi_0 \omega 
 \Biggl[  16\pi e^{2\alpha} 
{(\Omega-\omega)(\varepsilon+p)\over 1-v^2} \nonumber \\
 & & +{1 \over r}\omega,_r+{\cot\theta \over r^2}\omega,_\theta
  +\omega,_r(\beta+\nu),_r  \nonumber \\
&& +{1 \over r^2}\omega,_\theta(\beta + \nu),_\theta
   \Biggr] e^{2\beta-2\nu} r^3\sin^2\theta\, dr\, d\theta \Bigg\} ^{-1},
\label{GRV2_mod}
\end{eqnarray}

\begin{eqnarray}
\lambda_3' &\equiv&
4 \pi \int_0^{\pi} \int_0^{r_{max}}
   \left[ 3 p + (\varepsilon+p){v^2 \over (1-v^2)} 
  \right] e^{2\alpha+\beta}\, r^2 \, \nonumber \\
&& \sin\theta \,dr\, d\theta 
\times \Bigg\{
\int_0^{\pi} \int_0^{r_{max}} \Biggl[
-\nu\left( \partial^2\nu + \partial\nu\partial\beta \right) \nonumber \\
&&
+{1 \over 2}\alpha \left( \partial^2\beta + \partial\beta\partial\beta
                   \right)    -{1 \over 2r \tan\theta} 
    {\partial \alpha \over \partial r}
    \left(1 - e^{2\alpha-2\beta}\right) \nonumber \\
& &
   +{1\over 4r \tan\theta} {\partial \beta  \over \partial r}
     \left(1 - e^{2\alpha-2\beta}\right) \nonumber \\
& & 
     -{3\over 8}\omega 16\pi  e^{2\beta-2\nu}
    {(\varepsilon+p)(\Omega-\omega) \over (1-v^2)}  e^{2\beta-2\nu}
    r^2 \sin^2\theta 
\Biggr] \nonumber \\
&&  r^2e^{\beta}\sin\theta dr d\theta \Bigg\} ^{-1},
\label{GRV3_mod}
\end{eqnarray}
with another abridged notation
\begin{equation}
  \partial^2 \psi \equiv 
    {\partial^2 \psi \over {\partial r}^2} +{2 \over r}{\partial\psi \over \partial r}
  + {1\over r^2}{\partial^2 \psi \over {\partial \theta}^2}  
  + {1\over r^2\tan\theta}{\partial \psi \over \partial \theta} \ ,
\end{equation}
and $r_{max}$ is the distance of the truncated point beyond which actual 
numerical computations are not carried out in the KEH(OR) code.

This rewrite does not break the mathematical identity. In a sense,
it may make ``identity" more trivial, and then what information the 
identities provide us becomes unclear.
However, as far as the same expression of the identity in the same
code is used, they can play a role as indicators of accuracy
among models solved by each code.

\subsection{Tables of Models and Comparison} \label{s:tables}

The physical parameters of 45 models, computed in this comparison
project, are displayed in Tables~\ref{t:poly1} -- \ref{t:rot_real2}.  
All quantities are displayed
to three significant figures.  The lower and the upper bounds on each
quantity, as obtained by comparing results from the three codes, are
shown in the upper and lower rows for each corresponding quantity,
respectively. It follows that quantities expressed in a single row can
be regarded as "exact", to three significant figures.

Tables~\ref{t:poly1} and \ref{t:poly2} display results for polytropes with 
index $N = 0.5, \ 0.75, \ 1.0$
and $1.5$. For each value of the polytropic index $N$ we compute the
following models:
\begin{enumerate}
\item  a spherical Newtonian model (denoted by the symbol {\it sn}),
\item  a rapidly rotating Newtonian model ({\it rn}), 
\item  a nearly spherical, relativistic model ({\it sr}) , 
\item  the maximum mass model ({\it mr}) and 
\item a rapidly rotating relativistic model ({\it rr}).
\end{enumerate}

For the constant density case ($N = 0$), the spherical Newtonian and
spherical relativistic models are displayed in Table~\ref{t:spher_const}
and the rapidly
rotating Newtonian and rapidly rotating relativistic models in
Table~\ref{t:rot_const}. 
While all other models are specified by the central energy density
$\varepsilon_c$ and the ratio of the polar radius to the equatorial
radius $r_p/r_e$, constant density models are specified by their
central pressure and $r_p/r_e$.
 
For realistic equations of state, spherical models are shown in
Table~\ref{t:spher_real} 
and rotating models with $1.4 M_{\odot}$ ({\it br}: binary
pulsar mass and relativistic) and maximum mass models are shown in
Tables~\ref{t:rot_real1} and \ref{t:rot_real2}.  
For the equation of state L, model L(L)$mr$ uses four-point
Lagrange interpolation, while L(H)$mr$ is the same model but computed
using cubic Hermite interpolation.

From the tables displaying polytropic models 
one can see that the three codes have a good
agreement on most quantities especially for soft polytropes.
For stiff polytropes ($N<1.0$) the agreement is somewhat smaller.
For constant density models, the relative differences between the three
codes become several percent.
More sensitive quantities are the three redshifts and the eccentricity. 
It should be noted that redshift factors are local quantities which reflect the
metric potentials at each point. This implies that local values of the
metric potentials do not have the same agreement between different
numerical codes, as integrated global quantities.  For the
eccentricity, one needs to compute the length along the surface of the
star (see the definition of the eccentricity~(\ref{eccen})). Since in the
KEH codes $\mu=\cos \theta$ is used as the angular variable there arise 
numerical errors near the pole region, i.e. $\theta \simeq 0$.  Thus, the
differences in the values of the eccentricity also reflect this numerical
error due to the choice of coordinates. This causes differences of up to a
few percent in the eccentricity for rapidly rotating models.
On the other hand, global quantities such as angular velocity, mass, radius and 
angular momentum agree quite well among results of different codes.

From Tables~\ref{t:spher_real} -- \ref{t:rot_real2}, 
similar tendencies can be observed for realistic
equations of state. Models for the most  EOSs, except EOS CLES, have a good 
agreement between the three codes, although the agreement is not as good
as for polytropic models.. By comparing models constructed with 
EOSs WFF(FPS) and  WFF(NV), it is evident that the choice of the low 
density EOSs affects very little the structure of the star. 

The main reason for the large differences in the constant density case is that
the discontinuous density distribution is creating  Gibbs phenomena near 
the surface and this affects  all three codes.  The reason for the smaller agreement
for realistic EOSs, compared to polytropic EOSs, is that
the necessary interpolation between tabulated data affects the accuracy with 
which the equation of hydrostationary equilibrium is satisfied. For EOS L, the
choice of the interpolation scheme also affects the accuracy of the computed
models, with the cubic Hermite scheme being a better choice compared to
a four-point Lagrange interpolation (see the discussion in a later section).  For
the other realistic EOSs the choice of the interpolation scheme had a
negligible effect on the accuracy of computed models.

\begin{table*}
\caption{\label{t:det_poly1} Detailed comparison of polytropic models}
\begin{tabular}{ccccccc}
\hline
Model   & KEH(OR)  & KEH(SF) & BGSM  & diff1  & diff2   & diff3 \\
\hline
Model                      & N15sn           &                &                &       &       &       \\
$\bar \varepsilon_{\rm c}$ & 1.0000000e-09  & 1.0000000e-09  & 1.0000000e-09  &   &   &  \\
$r_p/r_e$                  & 9.6000000e-01  & 9.6000000e-01  & 9.6000000e-01  &  &  &  \\
$\bar \Omega$              & 6.5347464e-06  & 6.5349925e-06  & 6.5346091e-06  & -2e-5 & -4e-5 & 6e-5  \\
$\bar M$                   & 9.7631159e-05  & 9.7633565e-05  & 9.7634578e-05  & 4e-5  & -3e-5 & -1e-5 \\
$\bar M_0$                 & 9.7631237e-05  & 9.7633644e-05  & 9.7634657e-05  & 4e-5  & -3e-5 & -1e-5 \\
$\bar R_{circ}$            & 5.2907355e+01  & 5.2907099e+01  & 5.2908476e+01  & 2e-5  & 5e-6  & -3e-5 \\
$\bar J$                   & 3.5985713e-07  & 3.5988285e-07  & 3.5988120e-07  & 7e-5  & -7e-5 & 5e-6  \\
$\bar I$                   & 5.5068262e-02  & 5.5070123e-02  & 5.5073103e-02  & 9e-5  & -3e-5 & -5e-5 \\
$T/W$                      & 7.4801796e-03  & 7.4805723e-03  & 7.4802183e-03  & 5e-6  & -5e-5 & 5e-5  \\
$Z_p$                      & 1.9104721e-06  & 1.9105286e-06  & 1.9104977e-06  & 1e-5  & -3e-5 & 2e-5  \\
$Z^f_{eq}$                 & -3.4382698e-04 & -3.4383282e-04 & -3.4382701e-04 & 9e-8  & -2e-5 & 2e-5  \\
$Z^b_{eq}$                 & 3.4764792e-04  & 3.4765387e-04  &  3.4764801e-4  & 3e-7  & -2e-5 & 2e-5  \\
$e$                        & 2.7990046e-01  & 2.8386885e-01  & 2.8014870e-01  & 9e-4  & -1e-2 & 1e-2  \\
GRV2                       & 1.0578017e-04  & 6.3653021e-05  & -9.0173678e-08 &       &       &       \\
GRV3                       & -1.3712946e-05 & 7.8924860e-05  & -2.2726769e-07 &       &       &       \\
\hline
Model                      & N15mr           &                &                &       &       &       \\
$\bar \varepsilon_{\rm c}$ & 6.1000000e-02  & 6.1000000e-02  & 6.1000000e-02  & &  &  \\
$r_p/r_e$                  & 6.2000000e-01  & 6.2000000e-01  & 6.2000000e-01  &   &  &   \\
$\bar \Omega$              & 1.1082917e-01  & 1.1080248e-01  & 1.1079616e-01  & -3e-4 & 2e-4  & 6e-5  \\
$\bar M$                   & 2.9099968e-01  & 2.9091548e-01  & 2.9091876e-01  & -3e-4 & 3e-4  & -1e-5 \\
$\bar M_0$                 & 3.0433314e-01  & 3.0436572e-01  & 3.0437123e-01  & 1e-4  & -1e-4 & -2e-5 \\
$\bar R_{circ}$            & 2.8538213e+00  & 2.8537335e+00  & 2.8539175e+00  & 3e-5  & 3e-5  & -6e-5 \\
$\bar J$                   & 3.8794961e-02  & 3.8804941e-02  & 3.8806669e-02  & 3e-4  & -3e-4 & -5e-5 \\
$\bar I$                   & 3.5004286e-01  & 3.5021727e-01  & 3.5025283e-01  & 6e-4  & -5e-4 & -1e-4 \\
$T/W$                      & 4.7633988e-02  & 4.7508685e-02  & 4.7507826e-02  & -3e-3 & 3e-3  & 2e-5  \\
$Z_p$                      & 1.9429844e-01  & 1.9478524e-01  & 1.9478308e-01  & 3e-3  & -3e-3 & 1e-5  \\
$Z^f_{eq}$                 & -2.2475260e-01 & -2.2447647e-01 & -2.2448160e-01 & -1e-3 & 1e-3  & -2e-5 \\
$Z^b_{eq}$                 & 6.2263043e-01  & 6.2336170e-01  & 6.2335930e-01  & 1e-3  & -1e-3 & 4e-6  \\
$e$                        & 7.2977397e-01  & 7.3040218e-01  & 7.3099356e-01  & 2e-3  & -9e-4 & -8e-4 \\
GRV2                       & 3.8566454e-04  & 3.8935202e-03  & -4.5069715e-07 &       &       &       \\
GRV3                       & -5.3198333e-05 & 1.1087078e-04  & -2.5949035e-06 &       &       &       \\
\hline
Model                      & N05sn           &                &                &       &       &       \\
$\bar \varepsilon_{\rm c}$ & 1.0000000e-03  & 1.0000000e-03  & 1.0000000e-03  &   &   &   \\
$r_p/r_e$                  & 9.8000000e-01  & 9.8000000e-01  & 9.8000003e-01  &   &   &   \\
$\bar \Omega$              & 6.9205090e-03  & 6.9211147e-03  & 6.9078733e-03  & -2e-3 & -9e-5 & 2e-3  \\
$\bar M$                   & 6.2992939e-08  & 6.2995192e-08  & 6.2989922e-08  & -5e-5 & -4e-5 & 8e-5  \\
$\bar M_0$                 & 6.2993012e-08  & 6.2995264e-08  & 6.2989995e-08  & -5e-5 & -4e-5 & 8e-5  \\
$\bar R_{circ}$            & 3.0433836e-02  & 3.0433775e-02  & 3.0432960e-02  & -3e-5 & 2.e-6 & 3e-5  \\
$\bar J$                   & 1.3150934e-13  & 1.3153761e-13  & 1.3125388e-13  & -2e-3 & -2e-4 & 2e-3  \\
$\bar I$                   & 1.9002842e-11  & 1.9005263e-11  & 1.9000620e-11  & -1e-4 & -1e-4 & 2e-4  \\
$T/W$                      & 5.1994036e-03  & 5.2005153e-03  & 5.1792787e-03  & -3e-3 & -2e-4 & 4e-3  \\
$Z_p$                      & 2.0985378e-06  & 2.0986158e-06  & 2.0984061e-06  & -6e-5 & -3e-5 & 1e-4  \\
$Z^f_{eq}$                 & -2.0851997e-04 & -2.0851108e-04 & -2.0812950e-04 & -2e-3 & 4e-5  & 2e-3  \\
$Z^b_{eq}$                 & 2.1271705e-04  & 2.1270832e-04  & 2.1232632e-04  & -2e-3 & 4e-5  & 2e-3  \\
$e$                        & 1.9884365e-01  & 2.0484010e-01  & 1.9911770e-01  & 1e-3  & -3e-2 & 3e-2  \\
GRV2                       & -7.8558660e-06 & 2.5493102e-05  & -8.1720158e-06 &       &       &       \\
GRV3                       & -1.1614383e-05 & 9.1053508e-05  & -2.6069983e-06 &       &       &       \\
\hline
\end{tabular}
\end{table*}

\begin{table*}
\caption{\label{t:det_poly2} Detailed comparison of polytropic models (continued)}
\begin{tabular}{ccccccc}
\hline
Model   & KEH(OR)  & KEH(SF) & BGSM  & diff1  & diff2   & diff3 \\
\hline
Model                      & N05mr           &                &                &       &       &       \\
$\bar \varepsilon_{\rm c}$ & 1.0400000e+00  & 1.0400000e+00  & 1.0400000e+00  &  &   &   \\
$r_p/r_e$                  & 5.5000000e-01  & 5.5000000e-01  & 5.5000156e-01  &   &  &  \\
$\bar \Omega$              & 9.8004159e-01  & 9.7793068e-01  & 9.7787248e-01  & -2e-3 & 2e-3  & 6e-5  \\
$\bar M$                   & 1.5376721e-01  & 1.5302143e-01  & 1.5301513e-01  & -5e-3 & 5e-3  & 4e-5  \\
$\bar M_0$                 & 1.8250033e-01  & 1.8251809e-01  & 1.8250776e-01  & 4e-5  & -1e-4 & 6e-5  \\
$\bar R_{circ}$            & 5.3491240e-01  & 5.3492344e-01  & 5.3489664e-01  & -3e-5 & -2e-5 & 5e-5  \\
$\bar J$                   & 1.7214828e-02  & 1.7214474e-02  & 1.7217664e-02  & 2e-4  & 2e-5  & -2e-4 \\
$\bar I$                   & 1.7565405e-02  & 1.7602959e-02  & 1.7607269e-02  & 2e-3  & -2e-3 & -2e-4 \\
$T/W$                      & 1.4949334e-01  & 1.4721775e-01  & 1.4726559e-01  & -2e-2 & 2e-2  & -3e-4 \\
$Z_p$                      & 9.6404034e-01  & 9.7533189e-01  & 9.7519945e-01  & 1e-2  & -1e-2 & 1e-4  \\
$Z^f_{eq}$                 & -3.6445066e-01 & -3.6225810e-01 & -3.6208301e-01 & -7e-3 & 6e-3  & 4e-4  \\
$Z^b_{eq}$                 & 2.9851556e+00  & 3.0239590e+00  & 3.0243455e+00  & 1e-2  & -1e-2 & -1e-4 \\
$e$                        & 6.9783608e-01  & 6.9495486e-01  & 7.0174527e-01  & 6e-3  & 4e-3  & -1e-2 \\
GRV2                       & -2.9470467e-03 & 3.5351133e-02  & -5.3131831e-05 &       &       &       \\
GRV3                       & -7.2928314e-02 & 1.1781519e-04  & 1.3585788e-04  &       &       &       \\
\hline
\end{tabular}
\end{table*}

%%%%%%%%%%%%%%%%%%
%  const_detail.tex
%%%%%%%%%%%%%%%%%
\begin{table*}
\caption{\label{t:det_const1} Detailed comparison of constant density models} 
\begin{tabular}{ccccccc}
\hline
Model   & KEH(OR)  & KEH(SF) & BGSM  & diff1  & diff2   & diff3 \\
\hline
 Model                     & N00sn           &                &                &        &        &        \\
$\bar \varepsilon_{\rm c}$ &  1.0000000e+00 &  1.0000000e+00 &  1.0000000e+00 &        &        &        \\
$\bar p_{\rm c}$           &  1.0000000e-04 &  1.0000000e-04 &  1.0000000e-04 &        &        &        \\
$r_p/r_e$                  &  1.0000000e+00 &  1.0000000e+00 &  1.0000000e+00 &        &        &        \\
$\bar M$                   &  1.4049747e-06 &  1.4079464e-06 &  1.3811478e-06 &  -2e-2 &  -2e-3 &   2e-2 \\
$\bar M_0$                 &  1.4051472e-06 &  1.4081229e-06 &  1.3813135e-06 &  -2e-2 &  -2e-3 &   2e-2 \\
$\bar R_{circ}$            &  6.9250048e-03 &  6.9085570e-03 &  6.9085014e-03 &  -2e-3 &   2e-3 &   8e-6 \\
$Z_p$                      &  2.0197813e-04 &  2.0386139e-04 &  1.9998000e-04 &  -1e-2 &  -9e-3 &   2e-2 \\
$Z_{eq}^f$                 &  2.0197813e-04 &  2.0386139e-04 &  1.9998000e-04 &  -1e-2 &  -9e-3 &   2e-2 \\
$Z_{eq}^b$                 &  2.0197813e-04 &  2.0386139e-04 &  1.9998000e-04 &  -1e-2 &  -9e-3 &   2e-2 \\
GRV2                       &  8.5024396e-03 &  2.5005801e-02 &  1.9992785e-10 &        &        &        \\
GRV3                       &  1.1669659e-02 &  3.1034530e-02 &  2.4994651e-10 &        &        &        \\
\hline
 Model                     &  N00rn          &                &                &        &        &        \\ 
$\bar \varepsilon_{\rm c}$ &  1.0000000e+00 &  1.0000000e+00 &  1.0000000e+00 &        &        &        \\
$\bar p_{\rm c}$           &  1.0000000e-04 &  1.0000000e-04 &  1.0000000e-04 &        &        &        \\
$r_p/r_e$                  &  6.5000000e-01 &  6.5000000e-01 &  6.4996883e-01 &  -5e-5 &        &   5e-5 \\
$\bar \Omega$              &  1.0038565e+00 &  1.0145815e+00 &  1.0154202e+00 &   1e-2 &  -1e-2 &  -8e-4 \\
$\bar M$                   &  1.8558447e+00 &  1.8466846e+00 &  1.8532465e+00 &  -1e-3 &   5e-3 &  -4e-3 \\
$\bar M$                   &  2.0497584e-06 &  2.0637739e-06 &  2.0569322e-06 &   4e-3 &  -7e-3 &   3e-3 \\
$\bar M_0$                 &  2.0500389e-06 &  2.0640573e-06 &  2.0572047e-06 &   4e-3 &  -7e-3 &   3e-3 \\
$\bar R_{circ}$            &  9.0583317e-03 &  9.1109677e-03 &  9.1096124e-03 &   6e-3 &  -6e-3 &   2e-4 \\
$\bar J$                   &  6.7616039e-11 &  6.9668544e-11 &  6.9165592e-11 &   2e-2 &  -3e-2 &   7e-3 \\
$\bar I$                   &  6.7356280e-11 &  6.8667275e-11 &  6.8115244e-11 &   1e-2 &  -2e-2 &   8e-3 \\
$T/W$                      &  1.0792359e-01 &  1.1089017e-01 &  1.1416842e-01 &   6e-2 &  -3e-2 &  -3e-2 \\
$Z_p$                      &  2.8331355e-04 &  2.8580524e-04 &  2.8505153e-04 &   6e-3 &  -9e-3 &   3e-3 \\
$Z_{eq}^f$                 & -8.8147112e-03 & -8.9620804e-03 & -8.9698802e-03 &   2e-2 &  -2e-2 &  -9e-4 \\
$Z_{eq}^b$                 &  9.3813731e-03 &  9.5337271e-03 &  9.5400451e-03 &   2e-2 &  -2e-2 &  -7e-4 \\
$e$                        &  7.5985797e-01 &  7.6083298e-01 &  7.5926925e-01 &  -8e-4 &  -1e-3 &   2e-3 \\
GRV2                       & -1.0486413e-03 &  4.0913020e-02 &  1.5297023e-03 &        &        &        \\
GRV3                       & -1.9636423e-03 &  6.3495351e-04 &  1.8732802e-03 &        &        &        \\
\hline
 Model                     & N00sr           &                &                &        &        &        \\
$\bar \varepsilon_{\rm c}$ &  1.0000000e+00 &  1.0000000e+00 &  1.0000000e+00 &        &        &        \\
$\bar p_{\rm c}$           &  1.0000000e+00 &  1.0000000e+00 &  1.0000000e+00 &        &        &        \\
$r_p/r_e$                  &  1.0000000e+00 &  1.0000000e+00 &  1.0000000e+00 &        &        &        \\
$\bar M$                   &  1.1404379e-01 &  1.1257086e-01 &  1.1220252e-01 &  -2e-2 &   1e-2 &   3e-3 \\
$\bar M_0$                 &  1.5912882e-01 &  1.6139149e-01 &  1.5914795e-01 &   1e-4 &  -1e-2 &   1e-2 \\
$\bar R_{circ}$            &  3.0003488e-01 &  2.9923839e-01 &  2.9920671e-01 &  -3e-3 &   3e-3 &   1e-4 \\
$Z_p$                      &  9.7107121e-01 &  1.0138916e+00 &  1.0000000e+00 &   3e-2 &  -4e-2 &   1e-2 \\
$Z_{eq}^f$                 &  9.7107121e-01 &  1.0138916e+00 &  1.0000000e+00 &   3e-2 &  -4e-2 &   1e-2 \\
$Z_{eq}^b$                 &  9.7107121e-01 &  1.0138916e+00 &  1.0000000e+00 &   3e-2 &  -4e-2 &   1e-2 \\
GRV2                       & -4.2458406e-03 &  8.0302649e-03 &  1.2600343e-10 &        &        &        \\
GRV3                       &  3.4088425e-02 &  7.3549407e-03 &  1.3790569e-10 &        &        &        \\
\hline
\end{tabular}
\end{table*}

\begin{table*}
\caption{\label{t:det_const2} Detailed comparison of constant density models (continued)} 
\begin{tabular}{ccccccc}
\hline
Model   & KEH(OR)  & KEH(SF) & BGSM  & diff1  & diff2   & diff3 \\
\hline
 Model                     & N00rr           &                &                &        &        &        \\
$\bar \varepsilon_{\rm c}$ &  1.0000000e+00 &  1.0000000e+00 &  1.0000000e+00 &        &        &        \\
$\bar p_{\rm c}$           &  1.0000000e+00 &  1.0000000e+00 &  1.0000000e+00 &        &        &        \\
$r_p/r_e$                  &  7.0000000e-01 &  7.0000000e-01 &  7.0075459e-01 &   1e-3 &        & -1e-3  \\
$\bar \Omega$              &  1.3961457e+00 &  1.4071531e+00 &  1.3980528e+00 &   1e-3 &  -8e-3 &   7e-3 \\
$\bar M$                   &  1.8323492e+00 &  1.8121685e+00 &  1.8158693e+00 &  -9e-3 &   1e-2 &  -2e-3 \\
$\bar M$                   &  1.3889820e-01 &  1.3605365e-01 &  1.3462398e-01 &  -3e-2 &   2e-2 &   1e-2 \\
$\bar M_0$                 &  1.8665283e-01 &  1.8693714e-01 &  1.8377945e-01 &  -2e-2 &  -2e-3 &   2e-2 \\
$\bar R_{circ}$            &  3.4513863e-01 &  3.4566026e-01 &  3.4455898e-01 &  -2e-3 &  -2e-3 &   3e-3 \\
$\bar J$                   &  1.3838952e-02 &  1.4064876e-02 &  1.3733153e-02 &  -8e-3 &  -2e-2 &   2e-2 \\
$\bar I$                   &  9.9122546e-03 &  9.9952703e-03 &  9.8230574e-03 &  -9e-3 &  -8e-3 &   2e-2 \\
$T/W$                      &  1.6825846e-01 &  1.6281418e-01 &  1.6338670e-01 &  -3e-2 &   3e-2 &  -4e-3 \\
$Z_p$                      &  1.6030994e+00 &  1.7071394e+00 &  1.6720522e+00 &   4e-2 &  -6e-2 &   2e-2 \\
$Z_{eq}^f$                 & -1.5970481e-01 & -1.5951673e-01 & -1.5541165e-01 &  -3e-2 &   1e-3 &   3e-2 \\
$Z_{eq}^b$                 &  9.4121600e+00 &  1.1342393e+01 &  1.0436972e+01 &   1e-1 &  -2e-1 &   9e-2 \\
$e$                        &  7.0676794e-01 &  7.1111349e-01 &  7.0812060e-01 &   2e-3 &  -6e-3 &   4e-3 \\
GRV2                       & -1.5336628e-02 &  4.5755715e-02 &  1.7034437e-03 &        &        &        \\
GRV3                       & -1.0930542e-01 &  9.5032627e-04 &  4.2345737e-03 &        &        &        \\
\hline
\end{tabular}
\end{table*}
%%%%%%%%%%%%%%%%%%%%%%%%%%%%%%%%%%%%%%%%%%%%%%%%%%%%%%%%%%%%%%%%%%%%%%%%%%%%

%%%%%%%%%%%%%%%%%%
%  real_detail.tex
%%%%%%%%%%%%%%%%%%

\begin{table*}
\caption{\label{t:det_real1} Detailed comparison of models with realistic EOSs}
\begin{tabular}{ccccccc}
\hline
Model   & KEH(OR)  & KEH(SF) & BGSM  & diff1  & diff2   & diff3 \\
\hline
 Model                                & Gmr             &                &                &        &        &        \\
$\varepsilon_{\rm c} \ [g \ cm^{-3}]$ &  5.5828200e+15 &  5.5828200e+15 &  5.5828192e+15 &        &        &        \\
$r_p/r_e$                             &  5.7654000e-01 &  5.7654000e-01 &  5.7654388e-01 &        &        &        \\
$\Omega \ [ 10^4 \ s^{-1}]$           &  1.5778207e+00 &  1.5729910e+00 &  1.5731590e+00 &  -3e-3 &   3e-3 &  -1e-4 \\
$M \ [M_{\odot}]$                     &  1.5705842e+00 &  1.5670837e+00 &  1.5652708e+00 &  -3e-3 &   2e-3 &   1e-3 \\
$M_0 \ [M_{\odot}]$                   &  1.7915598e+00 &  1.7930714e+00 &  1.7913915e+00 &  -9e-5 &  -8e-4 &   9e-4 \\
$R_{circ} \ [km]$                     &  9.2371483e+00 &  9.2454157e+00 &  9.2471834e+00 &   1e-3 &  -9e-4 &  -2e-4 \\
$cJ/GM_{\odot}^2$                     &  1.5360080e+00 &  1.5378972e+00 &  1.5369015e+00 &   6e-4 &  -1e-3 &   7e-4 \\
$I \ [ 10^{45} g \ cm^2]$             &  8.5553062e-01 &  8.5923863e-01 &  8.5859068e-01 &   4e-3 &  -4e-3 &   8e-4 \\
$T/W$                                 &  1.0661337e-01 &  1.0541507e-01 &  1.0550611e-01 &  -1e-2 &   1e-2 &  -9e-4 \\
$Z_p$                                 &  7.4862286e-01 &  7.4468655e-01 &  7.4429886e-01 &  -6e-3 &   5e-3 &   5e-4 \\
$Z_{eq}^f$                            & -3.3808636e-01 & -3.3618145e-01 & -3.3645247e-01 &  -5e-3 &   6e-3 &  -8e-4 \\
$Z_{eq}^b$                            &  2.1006841e+00 &  2.1151171e+00 &  2.1140605e+00 &   6e-3 &  -7e-3 &   5e-4 \\
$e$                                   &  6.7374530e-01 &  6.8063333e-01 &  6.7954077e-01 &   9e-3 &  -1e-2 &   2e-3 \\
GRV2                                  &  1.3388534e-03 &  1.7877189e-02 & -3.6339736e-04 &        &        &        \\
GRV3                                  & -3.2386612e-02 &  9.7104987e-04 & -5.2521686e-04 &        &        &        \\
\hline
 Model                                & Lsr             &                &                &        &        &        \\
$\varepsilon_{\rm c} \ [g \ cm^{-3}]$ &  4.2995000e+14 &  4.2995000e+14 &  4.2995391e+14 &        &        &        \\
$r_p/r_e$                             &  1.0000000e+00 &  1.0000000e+00 &  1.0000000e+00 &        &        &        \\
$M \ [M_{\odot}]$                     &  1.4097969e+00 &  1.3884591e+00 &  1.4100000e+00 &   1e-4 &   2e-2 &  -2e-2 \\
$M_0 \ [M_{\odot}]$                   &  1.5235811e+00 &  1.4938211e+00 &  1.5253919e+00 &   1e-3 &   2e-2 &  -2e-2 \\
$R_{circ} \ [km]$                     &  1.4784119e+01 &  1.4819571e+01 &  1.4778437e+01 &  -4e-4 &  -2e-3 &   3e-3 \\
$Z_p$                                 &  1.7822863e-01 &  1.7557838e-01 &  1.7974175e-01 &   9e-3 &   2e-2 &  -2e-2 \\
$Z_{eq}^f$                            &  1.7829440e-01 &  1.7557838e-01 &  1.7974175e-01 &   8e-3 &   2e-2 &  -2e-2 \\
$Z_{eq}^b$                            &  1.7829440e-01 &  1.7557838e-01 &  1.7974175e-01 &   8e-3 &   2e-2 &  -2e-2 \\
GRV2                                  & -3.9884976e-03 &  1.7391300e-02 &  3.1884440e-03 &        &        &        \\
GRV3                                  &  2.4064437e-03 &  2.6180956e-02 &  1.2340895e-02 &        &        &        \\
\hline
 Model                                & L(L)mr          &                 &                &        &        &        \\ 
$\varepsilon_{\rm c} \ [g \ cm^{-3}]$ &  1.2022600e+15 &  1.2022600e+15  &  1.2022594e+15 &        &        &        \\
$r_p/r_e$                             &  5.5300000e-01 &  5.5300000e-01  &  5.5300318e-01 &        &        &        \\
$\Omega \ [ 10^4 \ s^{-1}]$           &  8.1797409e-01 &  8.1681214e-01  &  8.1674028e-01 &  -2e-3 &   1e-3 &   9e-5 \\
$M \ [M_{\odot}]$                     &  3.3254178e+00 &  3.3067743e+00  &  3.3156417e+00 &  -3e-3 &   6e-3 &  -3e-3 \\
$M_0 \ [M_{\odot}]$                   &  3.9164459e+00 &  3.8979008e+00  &  3.9209347e+00 &   1e-3 &   5e-3 &  -6e-3 \\
$R_{circ} \ [km]$                     &  1.8472119e+01 &  1.8460489e+01  &  1.8475482e+01 &   2e-4 &   6e-4 &  -8e-4 \\
$cJ/GM_{\odot}^2$                     &  7.7807280e+00 &  7.7201999e+00  &  7.8026429e+00 &   3e-3 &   8e-3 &  -1e-2 \\
$I \ [ 10^{45} g \ cm^2]$             &  8.3595026e+00 &  8.3065196e+00  &  8.3959624e+00 &   4e-3 &   6e-3 &  -1e-2 \\
$T/W$                                 &  1.3918072e-01 &  1.3669628e-01  &  1.3748748e-01 &  -1e-2 &   2e-2 &  -6e-3 \\
$Z_p$                                 &  8.2621901e-01 &  8.2779278e-01  &  8.3079056e-01 &   6e-3 &  -2e-3 &  -4e-3 \\
$Z_{eq}^f$                            & -3.4862304e-01 &  -3.4713361e-01 & -3.4699020e-01 &  -5e-3 &  4e-3  &  4e-4  \\
$Z_{eq}^b$                            &  2.4309406e+00 &  2.4463175e+00  &  2.4596968e+00 &   1e-2 &  -6e-3 &  -5e-3 \\
$e$                                   &  6.9748441e-01 &  7.0355618e-01  &  7.0273768e-01 &   8e-3 &  -9e-3 &   1e-3 \\
GRV2                                  &  9.4730085e-04 &  3.4941975e-02  & -2.2424315e-03 &        &        &        \\
GRV3                                  & -4.0789432e-02 &  2.4558001e-04  & -2.1163857e-03 &        &        &        \\
\hline
\end{tabular}
\end{table*}

\begin{table*}
\caption{\label{t:det_real2} Detailed comparison of models with realistic EOSs (continued)}
\begin{tabular}{ccccccc}
\hline
Model   & KEH(OR)  & KEH(SF) & BGSM  & diff1  & diff2   & diff3 \\
\hline
 Model                                & L(H)mr          &                &                &       &       &       \\
$\varepsilon_{\rm c} \ [g \ cm^{-3}]$ & 1.2432200e+15  & 1.2022600e+15  & 1.2020680e+15  & -3e-2 & 3e-2  & 2e-4  \\
$r_p/r_e$                             & 5.5521963e-01  & 5.5300000e-01  & 5.5423989e-01  & -2e-3 & 4e-3  & -2e-3 \\
$\Omega \ [ 10^4 \ s^{-1}]$           & 8.2083986e-01  & 8.1441391e-01  & 8.1328587e-01  & -9e-3 & 8e-3  & 1e-3  \\
$M \ [M_{\odot}]$                     & 3.3030613e+00  & 3.2951034e+00  & 3.2897723e+00  & -4e-3 & 2e-3  & 2e-3  \\
$M_0 \ [M_{\odot}]$                   & 3.8763017e+00  & 3.8823439e+00  & 3.8749019e+00  & -4e-4 & -2e-3 & 2e-3  \\
$R_{circ} \ [km]$                     & 1.8375548e+01  & 1.8488275e+01  & 1.8475487e+01  & 5e-3  & -6e-3 & 7e-4  \\
$cJ/GM_{\odot}^2$                     & 7.6167631e+00  & 7.6420920e+00  & 7.6199254e+00  & 4e-4  & -3e-3 & 3e-3  \\
$I \ [ 10^{45} g \ cm^2]$             & 8.1547703e+00  & 8.2466926e+00  & 8.2341775e+00  & 1e-2  & -1e-2 & 2e-3  \\
$T/W$                                 & 1.3741062e-01  & 1.3597594e-01  & 1.3602745e-01  & -1e-2 & 1e-2  & -4e-4 \\
$Z_p$                                 & 8.1941726e-01  & 8.2094860e-01  & 8.1829405e-01  & -1e-3 & -2e-3 & 3e-3  \\
$Z^f_{eq}$                            & -3.4820132e-01 & -3.4697590e-01 & -3.4552738e-01 & -8e-3 & 4e-3  & 4e-3  \\
$Z^b_{eq}$                            & 2.4176003e+00  & 2.4208889e+00  & 2.4115282e+00  & -3e-3 & -1e-3 & 4e-3  \\
$e$                                   & 6.9758545e-01  & 7.0330533e-01  & 7.0159644e-01  & 6e-3  & -8e-3 & 2e-3  \\
GRV2                                  & -2.2119050e-03 & 3.0670163e-02  & 6.9885050e-06  &       &       &       \\
GRV3                                  & -5.3441095e-02 & 7.8090726e-04  & 1.1634685e-04  &       &       &       \\
\hline
 Model                                & WFF(FPS)sr       &                &                &        &        &        \\
$\varepsilon_{\rm c} \ [g \ cm^{-3}]$ &  1.2161500e+15 &  1.2161500e+15 &  1.2161501e+15 &        &        &        \\
$r_p/r_e$                             &  1.0000000e+00 &  1.0000000e+00 &  1.0000000e+00 &        &        &        \\
$M \ [M_{\odot}]$                     &  1.4139823e+00 &  1.4097068e+00 &  1.4100000e+00 &  -3e-3 &   3e-3 &  -2e-4 \\
$M_0 \ [M_{\odot}]$                   &  1.5740517e+00 &  1.5720031e+00 &  1.5725139e+00 &  -1e-3 &   1e-3 &  -3e-4 \\
$R_{circ} \ [km]$                     &  1.0917706e+01 &  1.0916388e+01 &  1.0916799e+01 &  -8e-5 &   1e-4 &  -4e-5 \\
$Z_p$                                 &  2.6875136e-01 &  2.7106271e-01 &  2.7114022e-01 &   9e-3 &  -9e-3 &  -3e-4 \\
$Z_{eq}^f$                            &  2.6889160e-01 &  2.7106271e-01 &  2.7114022e-01 &   8e-3 &  -8e-3 &  -3e-4 \\
$Z_{eq}^b$                            &  2.6889160e-01 &  2.7106271e-01 &  2.7114022e-01 &   8e-3 &  -8e-3 &  -3e-4 \\
GRV2                                  & -9.5750969e-04 &  4.0179178e-04 &  4.4407986e-05 &        &        &        \\
GRV3                                  &  1.6329150e-02 &  4.9392946e-04 &  7.0678176e-05 &        &        &        \\
\hline
 Model                                & WFF(FPS)br       &                &                &        &        &        \\
$\varepsilon_{\rm c} \ [g \ cm^{-3}]$ &  9.7000000e+14 &  9.7000000e+14 &  9.7000000e+14 &        &        &        \\
$r_p/r_e$                             &  6.3000000e-01 &  6.3000000e-01 &  6.3000000e-01 &        &        &        \\
$\Omega \ [ 10^4 \ s^{-1}]$           &  7.0057052e-01 &  7.0002713e-01 &  7.0004420e-01 &  -8e-4 &   8e-4 &  -2e-5 \\
$M \ [M_{\odot}]$                     &  1.4166605e+00 &  1.4159575e+00 &  1.4151900e+00 &  -1e-3 &   5e-4 &   5e-4 \\
$M_0 \ [M_{\odot}]$                   &  1.5461089e+00 &  1.5466864e+00 &  1.5458039e+00 &  -2e-4 &  -4e-4 &   6e-4 \\
$R_{circ} \ [km]$                     &  1.4253215e+01 &  1.4252524e+01 &  1.4255153e+01 &   1e-4 &   5e-5 &  -2e-4 \\
$cJ/GM_{\odot}^2$                     &  1.3113239e+00 &  1.3120877e+00 &  1.3109603e+00 &  -3e-4 &  -6e-4 &   9e-4 \\
$I \ [ 10^{45} g \ cm^2]$             &  1.6449696e+00 &  1.6472547e+00 &  1.6457991e+00 &   5e-4 &  -1e-3 &   9e-4 \\
$T/W$                                 &  1.0155753e-01 &  1.0094356e-01 &  1.0095316e-01 &  -6e-3 &   6e-3 &  -1e-4 \\
$Z_p$                                 &  2.8235936e-01 &  2.8351010e-01 &  2.8339205e-01 &   4e-3 &  -4e-3 &   4e-4 \\
$Z_{eq}^f$                            & -2.1053187e-01 & -2.0979686e-01 & -2.0993514e-01 &  -3e-3 &   4e-3 &  -7e-4 \\
$Z_{eq}^b$                            &  8.1178465e-01 &  8.1358461e-01 &  8.1343105e-01 &   2e-3 &  -2e-3 &   2e-4 \\
$e$                                   &  7.1314329e-01 &  7.1756280e-01 &  7.1517800e-01 &   3e-3 &  -6e-3 &   3e-3 \\
GRV2                                  & -6.0505587e-04 &  2.6460925e-02 & -2.3145939e-05 &        &        &        \\
GRV3                                  &  2.1411844e-03 &  3.8377908e-04 & -3.7280705e-05 &        &        &        \\
\hline
\end{tabular}
\end{table*}

\begin{table*}
\caption{\label{t:det_real3} Detailed comparison of models with realistic EOSs (continued)}
\begin{tabular}{ccccccc}
\hline
Model   & KEH(OR)  & KEH(SF) & BGSM  & diff1  & diff2   & diff3 \\
\hline
 Model                                & CLESmr          &                &                &        &        &        \\
$\varepsilon_{\rm c} \ [g \ cm^{-3}]$ &  4.2000000e+14 &  4.2000000e+14 &  4.1928573e+14 &        &        &        \\
$r_p/r_e$                             &  5.3085938e-01 &  5.3085938e-01 &  5.3150938e-01 &        &        &        \\
$\Omega \ [ 10^4 \ s^{-1}]$           &  6.0622590e-01 &  6.0321091e-01 &  6.0275394e-01 &  -6e-3 &   5e-3 &   8e-4 \\
$M \ [M_{\odot}]$                     &  6.7113216e+00 &  6.6627293e+00 &  6.6394450e+00 &  -1e-2 &   7e-3 &   4e-3 \\
$M_0 \ [M_{\odot}]$                   &  8.3688660e+00 &  8.3951155e+00 &  8.3562334e+00 &  -2e-3 &  -3e-3 &   5e-3 \\
$R_{circ} \ [km]$                     &  2.8623928e+01 &  2.8660015e+01 &  2.8649929e+01 &   9e-4 &  -1e-3 &   3e-4 \\
$cJ/GM_{\odot}^2$                     &  3.6211896e+01 &  3.6395860e+01 &  3.6124793e+01 &  -2e-3 &  -5e-3 &   8e-3 \\
$I \ [ 10^{45} g \ cm^2]$             &  5.2494829e+01 &  5.3026814e+01 &  5.2671785e+01 &   3e-3 &  -1e-2 &   7e-3 \\
$T/W$                                 &  1.9699723e-01 &  1.9169339e-01 &  1.9176893e-01 &  -3e-2 &   3e-2 &  -4e-4 \\
$Z_p$                                 &  1.4566262e+00 &  1.4908621e+00 &  1.4820382e+00 &   2e-2 &  -2e-2 &   6e-3 \\
$Z_{eq}^f$                            & -4.0695375e-01 & -4.0247884e-01 & -4.0233644e-01 &  -1e-2 &   1e-2 &   4e-4 \\
$Z_{eq}^b$                            &  5.8899056e+00 &  6.1402011e+00 &  6.0575089e+00 &   3e-2 &  -4e-2 &   1e-2 \\
$e$                                   &  7.1993250e-01 &  7.2912408e-01 &  7.2745777e-01 &   1e-2 &  -1e-2 &   2e-3 \\
GRV2                                  & -7.4679128e-03 &  7.7678310e-02 &  5.8321139e-04 &        &        &        \\
GRV3                                  & -1.5019570e-01 &  9.3185944e-04 &  8.1316752e-04 &        &        &        \\
\hline
\end{tabular}
\end{table*}

\subsection{Detailed comparison}

In order to investigate further the differences among numerical results
obtained by the three codes, we show more detailed results
for models: N15$sn$, N15$mr$ and N05$sn$ in Table~\ref{t:det_poly1},
N05$mr$ in Table~\ref{t:det_poly2}, N00$sn$, 
N00$rn$ and  N00$sr$ in Table~\ref{t:det_const1}, N00$rr$ in 
Table~\ref{t:det_const2}, G$mr$, L$sr$ and L(L)$mr$ in Table~\ref{t:det_real1},
L(H)$mr$, WFF(FPS)$sr$ and WFF(FPS)$br$ in Table~\ref{t:det_real2} and
CLES$mr$ in Table~\ref{t:det_real3}.  
In these tables, values to eight figures 
for each physical  quantity are shown, as well as the relative differences among results 
of the three codes.

The three relative differences ${\rm diff1}$, ${\rm diff2}$ and ${\rm diff3}$
are defined as
\begin{eqnarray}
  {\rm diff1} & \equiv & {{\rm BGSM}    - {\rm KEH(OR)} \over {\rm KEH(OR)}} \ , \\
  {\rm diff2} & \equiv & {{\rm KEH(OR)} - {\rm KEH(SF)} \over {\rm KEH(SF)}} \ , \\
  {\rm diff3} & \equiv & {{\rm KEH(SF)} - {\rm BGSM}    \over {\rm BGSM}}    \ .
\end{eqnarray}

From Tables~\ref{t:det_poly1} and \ref{t:det_poly2}, 
we see that the agreement between the KEH(SF) and BGSM codes
for the rapidly rotating, relativistic polytropic models models N15$mr$ and
N05$mr$ is between $10^{-4}$ and $3 \times 10^{-4}$ in all computed quantities
(global and local), except for the eccentricity (due to reasons we explained in the
previous section). This very good agreement shows that both the BGSM and KEH
schemes (the latter when applying boundary conditions exactly at infinity) are
suitable for the construction of highly accurate initial-data configurations of
rapidly rotating relativistic stars,  modeled as polytropes with typical index
$N>0.5$.  

From the same tables, we see that the agreement between the KEH(OR) and BGSM
codes is similar to the agreement between KEH(SF) and BGSM in the global
quantities of model N15$mr$ but to within $10^{-3}$ for the local quantities
of this model. For the stiffer polytrope N05$mr$ the agreement is  
$10^{-3}$ and $10^{-2}$ for global and local quantities, respectively. This 
difference in accuracy between KEH(SF) and KEH(OR) is expected, since in KEH(OR) 
 boundary conditions are applied only approximately at a finite distance
from the star. Considering how close to the star the domain of integration is
truncated, the KEH(OR) code performs very well. This is explained as follows:

Since the integration is performed over only a finite region, the truncated 
part of the integral, $I_{tr}$,  can be expressed as
\begin{equation}
 I_{tr}({\bf r}) \equiv \int_{out} S({\bf r^{'}}) G({\bf r}, {\bf r^{'}}) 
d^3 {\bf r^{'}}, 
\end{equation}
where $S$ and $G$ are the source term and the Green's function, respectively,
and the subscript $out$ denotes that the integration covers the region
with $r \ge R_{max}$. Here $R_{max}$ is the maximum radius of the  
region of integration. Although the source terms of the integrals for 
the  metric potentials in 
the KEH(OR) scheme contain both matter and the metric terms,
only the metric terms contribute to the integral the outside of the star. 
Since we impose asymptotically flat conditions for the metric potentials,
the radial dependence of the source term can be considered to be
\begin{equation}
S \sim {1 \over r^k} ,
\end{equation}
where $k \ge 4$, because of the behavior of metric functions at
large radii.  Consequently, $I_{tr}$ can be roughly expressed as
\begin{eqnarray}
I_{tr}({\bf r}) & \sim & {\rm constant} ,   \qquad (r < r_e < R_{max})
                             \nonumber \\
                & \sim & {1 \over r^{k-2}} 
     \ln \left( 1-{r \over R_{max}} \right) .  \qquad (r_e < r < R_{max})
\end{eqnarray}
From the above equation, we can see the following:
First, for the stellar interior, i.e. $r < r_e$,
$I_{tr} \sim constant $, so that the relative error $I_{tr}/I_0$
can be very small where $I_0$ is the exact value of the integral.
Second, for the outer region of the stars, i.e. $r_e \le r \le R_{max}$,
the value of $I_{tr}$ becomes larger for larger values of $r$ because
of the logarithmic term. Third, for the same region, i.e.
$r_e \le r \le R_{max}$, if the source term depends weakly on
the radial coordinate, i.e., for smaller values of $k$, 
the contribution of the truncated part becomes larger. This occurs
for the potentials $\nu$ and $\omega$. However, for the metric
functions $B$ and $\zeta$, since values of $k$ are large, 
the relative differences are rather small. 

In the extreme case of constant density models (Tables~\ref{t:det_const1} and
\ref{t:det_const2}), the three codes
agree on the computed physical quantities typically only within a few percent and
this is caused by the sharp density discontinuity at the surface of the 
star. The numerical schemes in this comparison assume that the density 
distribution is a smooth function of coordinates, thus, in the case of
density discontinuities, this assumption is violated and Gibbs phenomena
appear, resulting in low accuracy of the computed models. 

From Tables~\ref{t:det_real1}--\ref{t:det_real3} it follows that the agreement 
of the KEH(SF) code to the BGSM
code is between $10^{-3}$ and $10^{-4}$ for realistic EOSs, except for
EOS CLES, where the agreement is an order of magnitude smaller. KEH(OR)
and BGSM agree on the realistic models within $10^{-2}$ and $10^{-3}$, i.e.
similar to the agreement for the $N=0.5$ polytrope. The somewhat lesser
agreement for realistic EOSs is due to the use of interpolation between
the tabulated equation of state data (see the discussion next).

In Tables~\ref{t:det_poly1} to \ref{t:det_real3}, 
we also display the virial quantities GRV2 and GRV3. In the ideal case,
these should exactly vanish, so the smaller the values for GRV2 and GRV3 are, the 
better is the accuracy of the computed model. The opposite is not always true,
i.e. in some models the computed values for GRV2 or GRV3 do not reflect an overall
better agreement in physical quantities among the different codes. This indicates that
the computation of GRV2 and GRV3 may itself be prone to numerical error. This seems
to be the case for GRV2 in rapidly rotating models computed with the KEH(SF) code, where
one first has to interpolate data between $\cos \theta$ and $\theta$ grids to be
able to compute GRV2. Note that the displayed values of the two virial quantities 
for the KEH(OR) code correspond to the modified virial identities  
(\ref{GRV2_mod}) and (\ref{GRV3_mod}) and not to the original identities. 
The computation of GRV3 for the KEH(OR) code is affected significantly by the
truncation of the domain of integration.

\begin{figure}
\resizebox{\hsize}{10.5cm}{\includegraphics{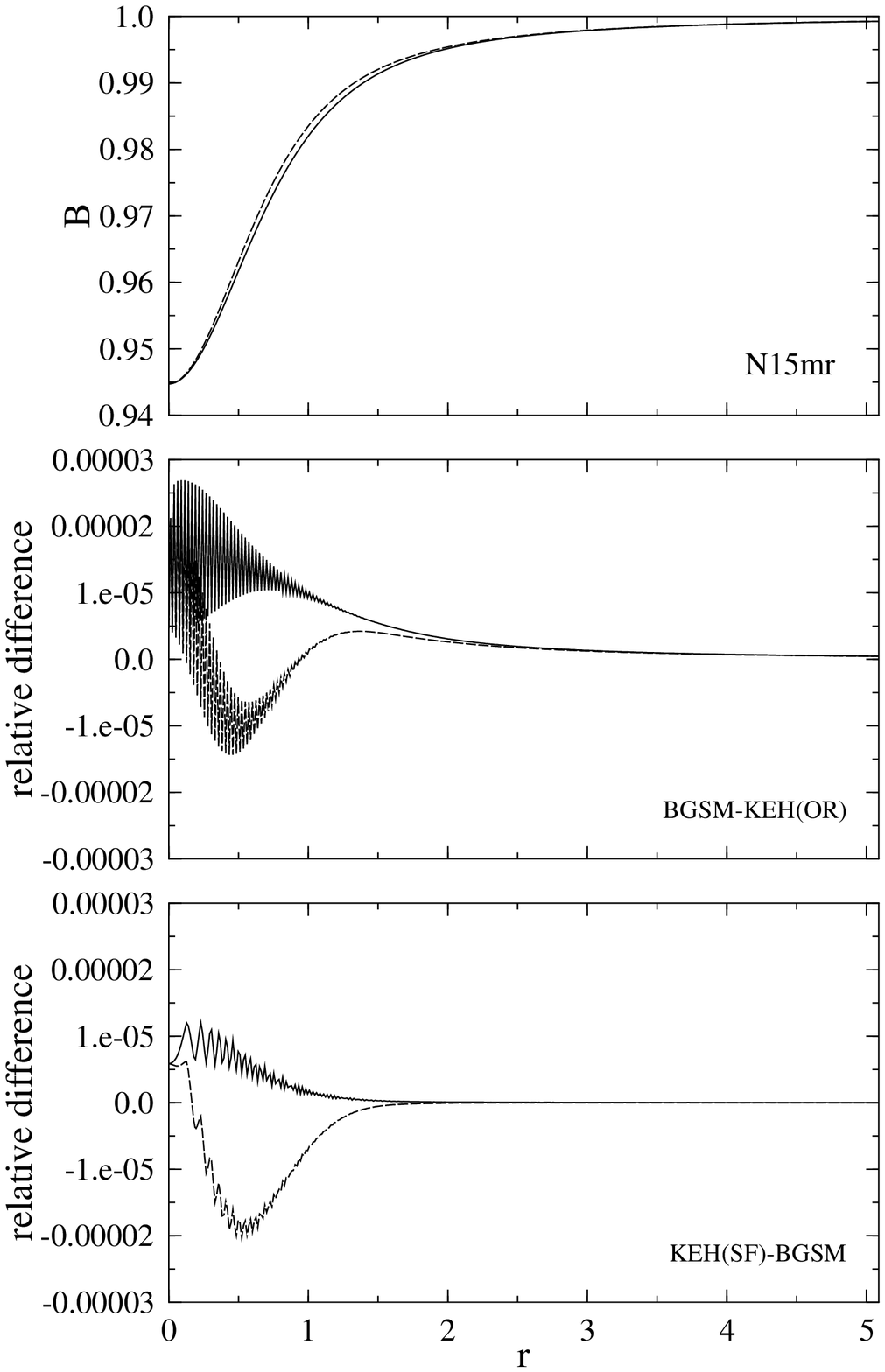}}
\caption{Metric potential $B$ as a function of coordinate radius for model N15$mr$ (upper 
panel). Relative difference of $B$ for the same model constructed with the BGSM and KEH(OR)
codes (middle panel) and with the KEH(SF) and BGSM codes (lower panel). The solid graph
corresponds to $\theta= \pi/2$ (equatorial plane) and the dashed line to $\theta=0$ (axis
of rotation). The largest value of $r$ displayed, corresponds to twice the coordinate 
radius of the surface of the star in the equatorial plane.}
\end{figure}

\begin{figure}[p]
\resizebox{\hsize}{10.5cm}{\includegraphics{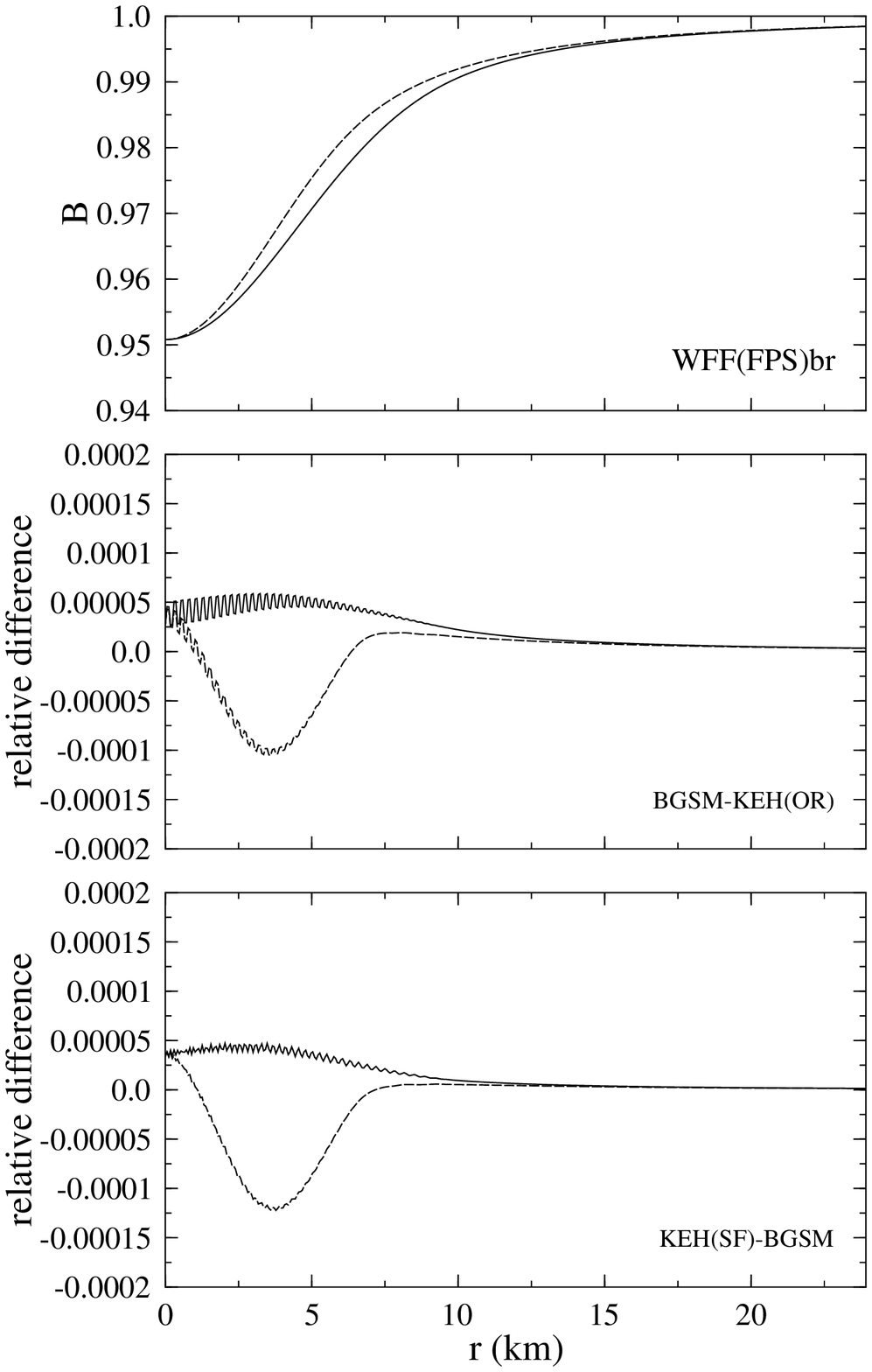}}
\caption{Same as Figure 1 but for the metric potential $B$ of model WFF(FPS)$br$.}
\end{figure}

\begin{figure}[p]
\resizebox{\hsize}{10.5cm}{\includegraphics{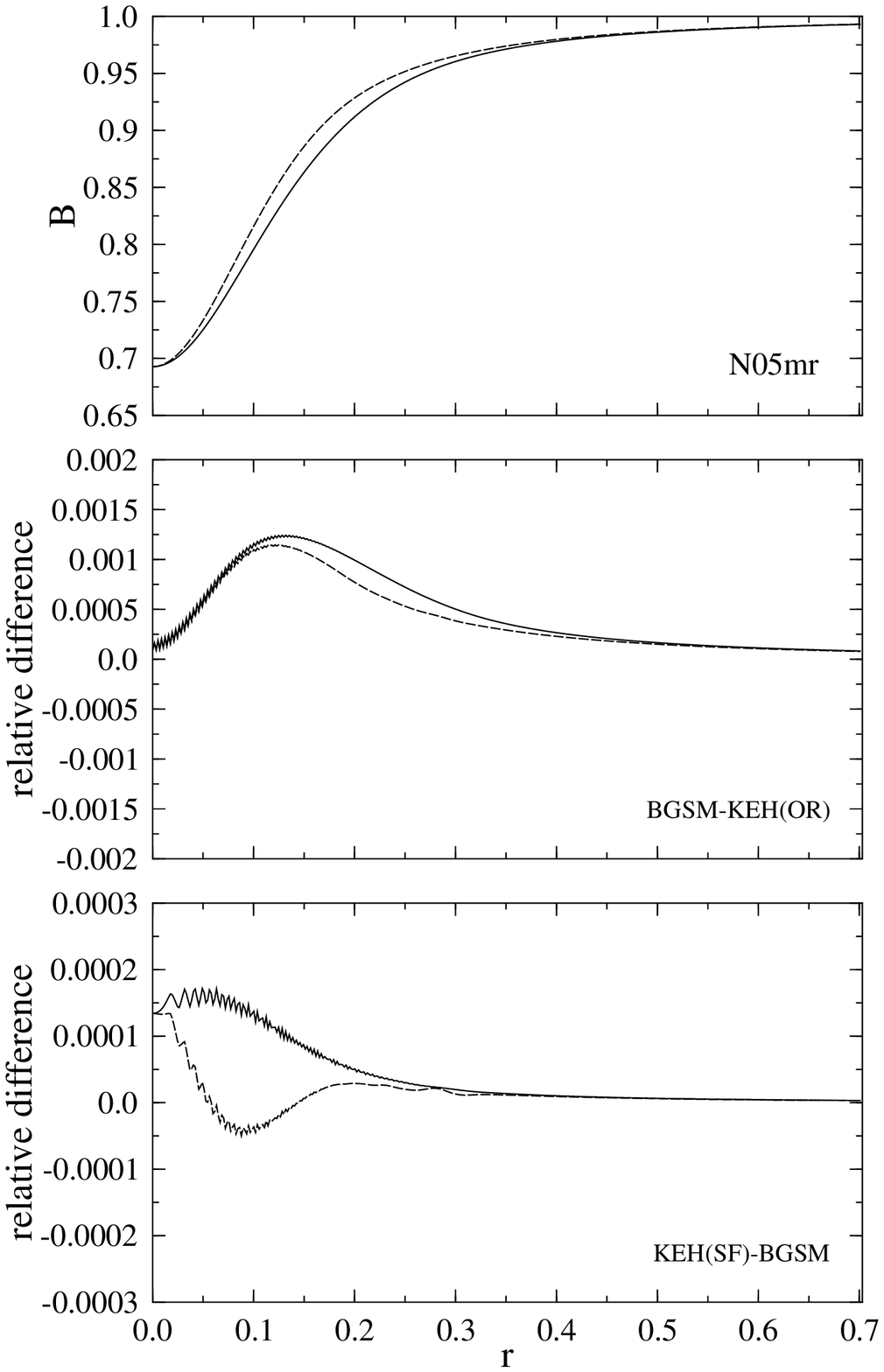}}
\caption{Same as Figure 1 but for the metric potential $B$ of model N05$mr$.}
\end{figure}

\begin{figure}[p]
\resizebox{\hsize}{10.5cm}{\includegraphics{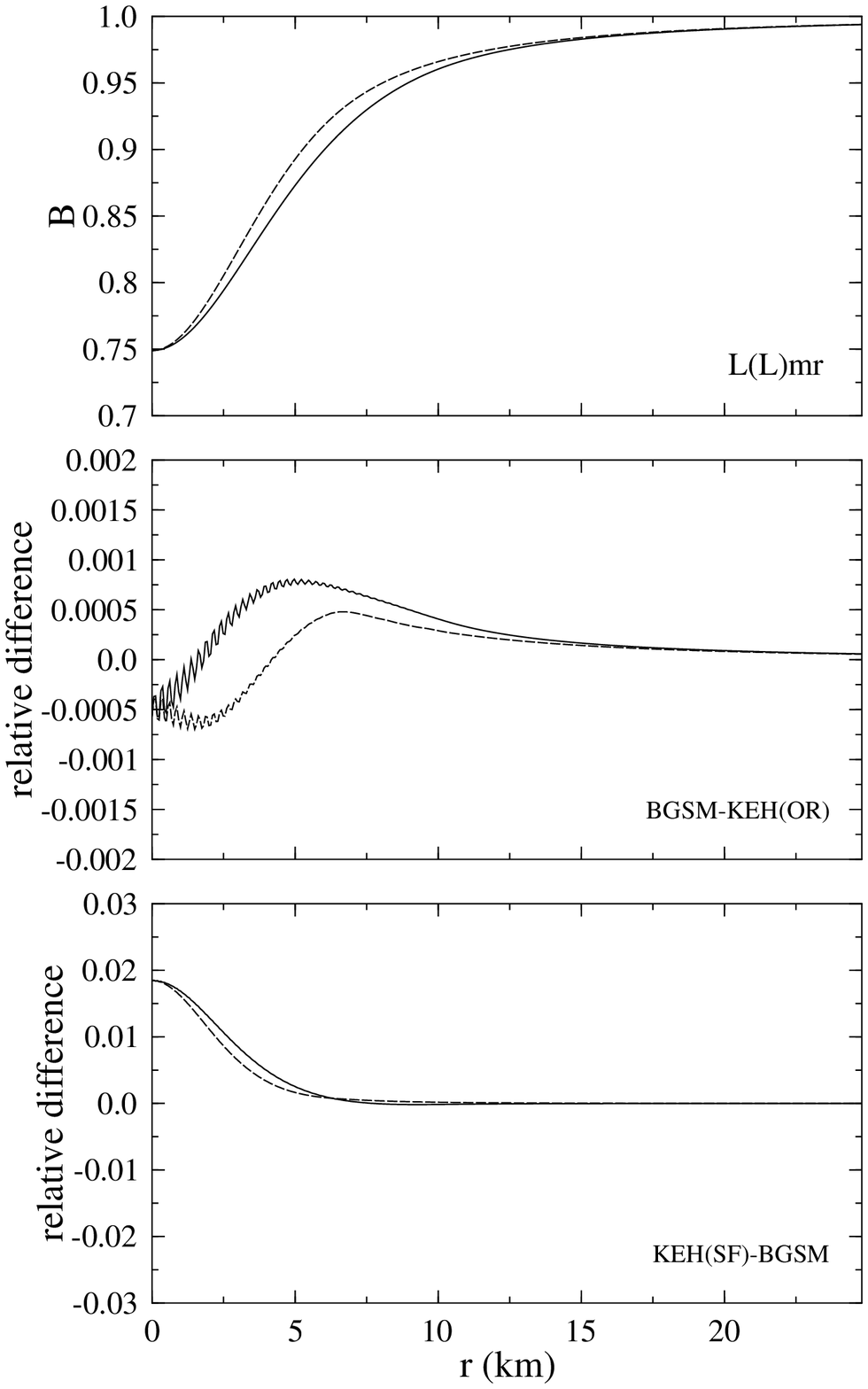}}
\caption{Same as Figure 1 but for the metric potential $B$ of model L(L)$mr$.}
\end{figure}

\begin{figure}[p]
\resizebox{\hsize}{10.5cm}{\includegraphics{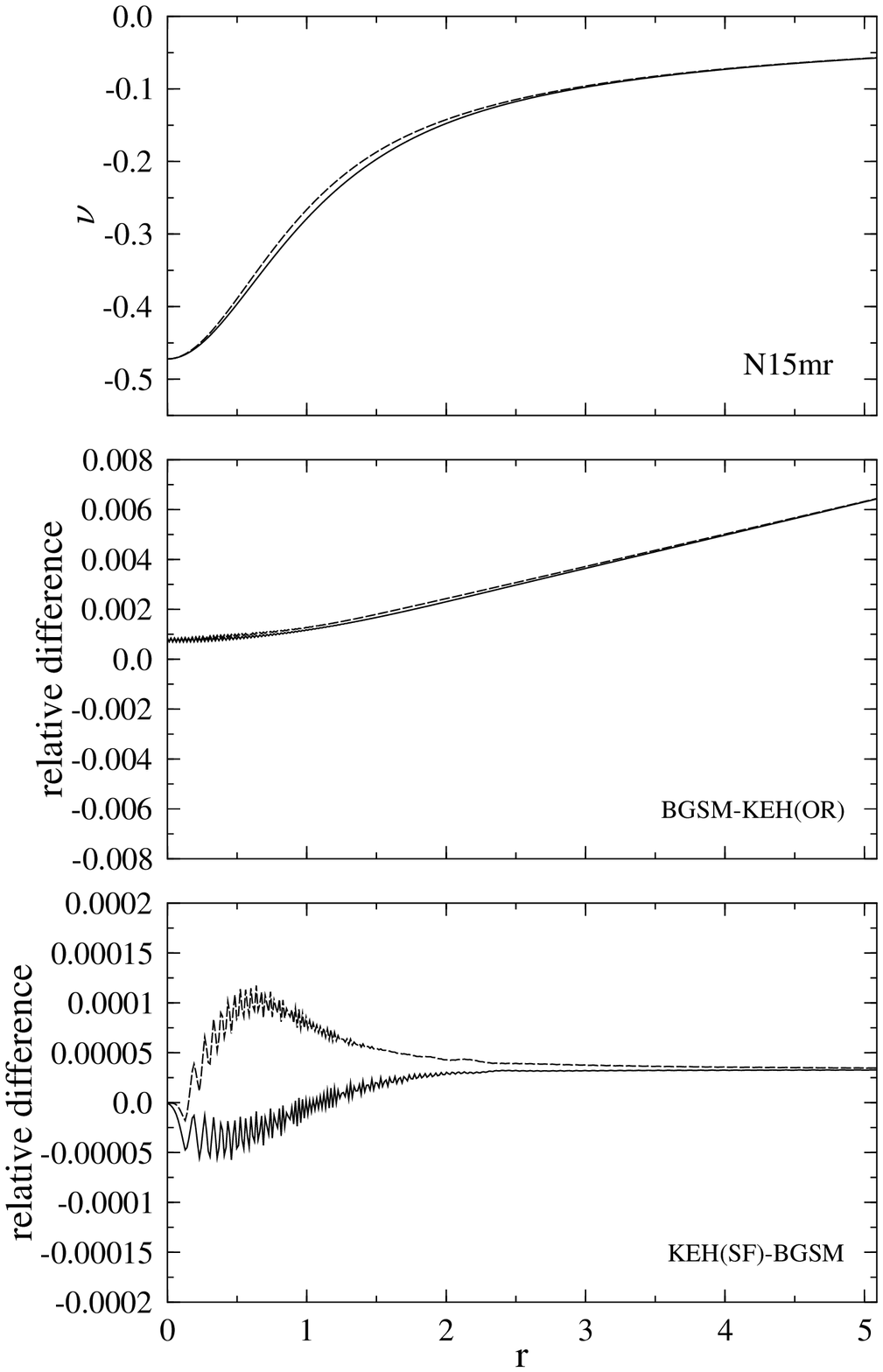}}
\caption{Same as Figure 1 but for the metric potential $\nu$ of model N15$mr$.}
\end{figure}

\begin{figure}
\resizebox{\hsize}{10.5cm}{\includegraphics{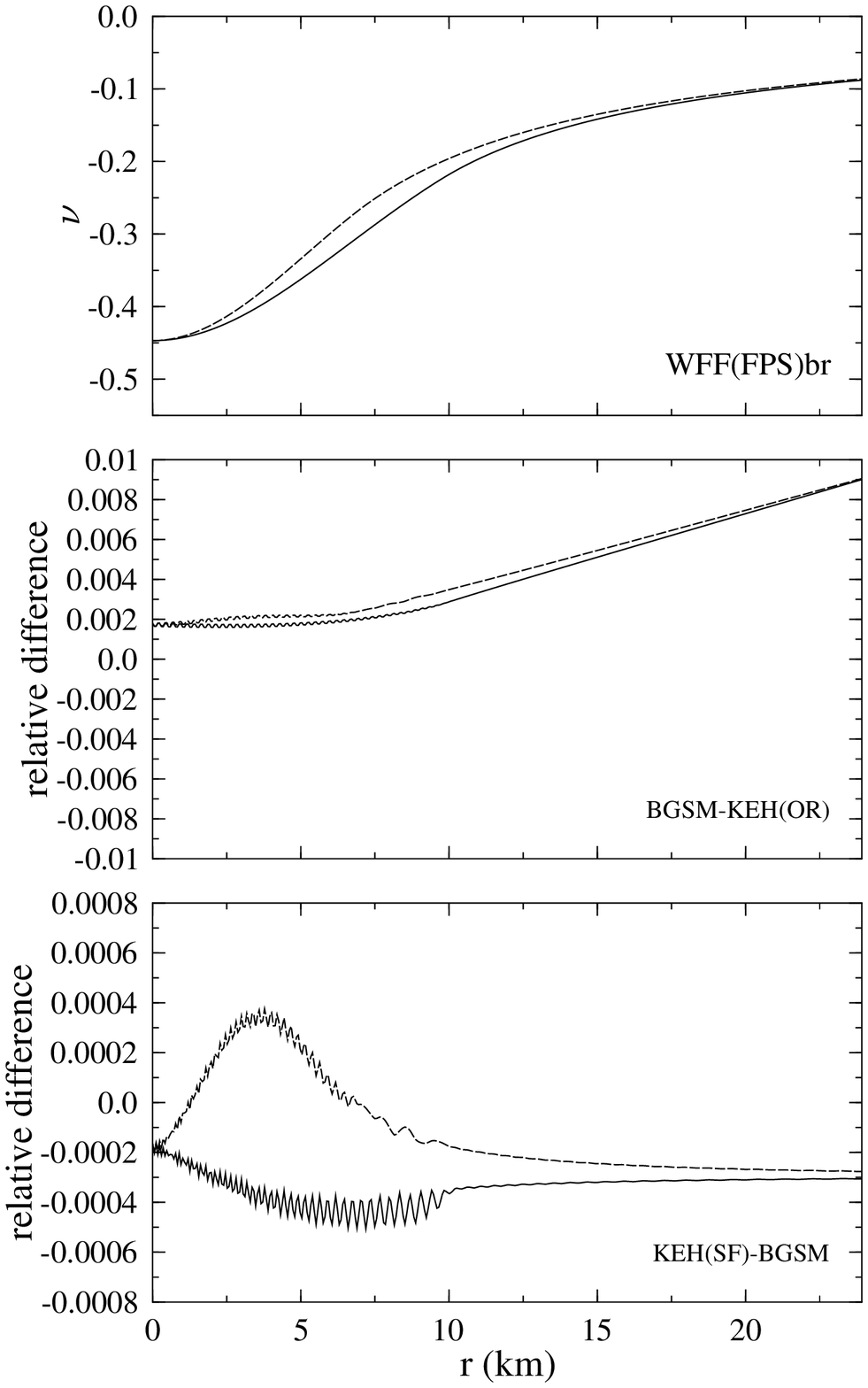}}
\caption{Same as Figure 1 but for the metric potential $\nu$ of model WFF(FPS)$br$.}
\end{figure}

\begin{figure}
\resizebox{\hsize}{10.5cm}{\includegraphics{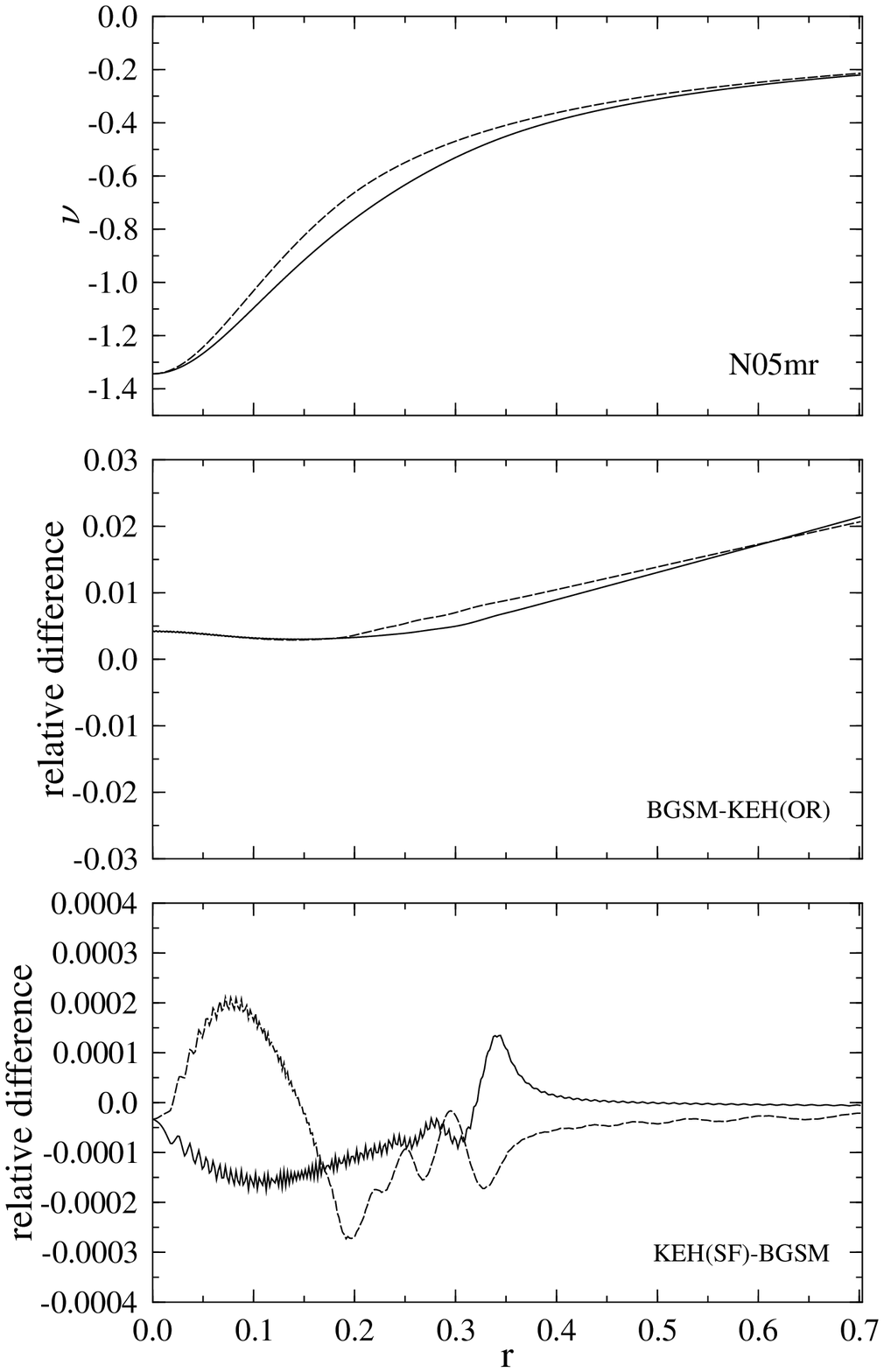}}
\caption{Same as Figure 1 but for the metric potential $\nu$ of model N05$mr$.}
\end{figure}

\begin{figure}
\resizebox{\hsize}{10.5cm}{\includegraphics{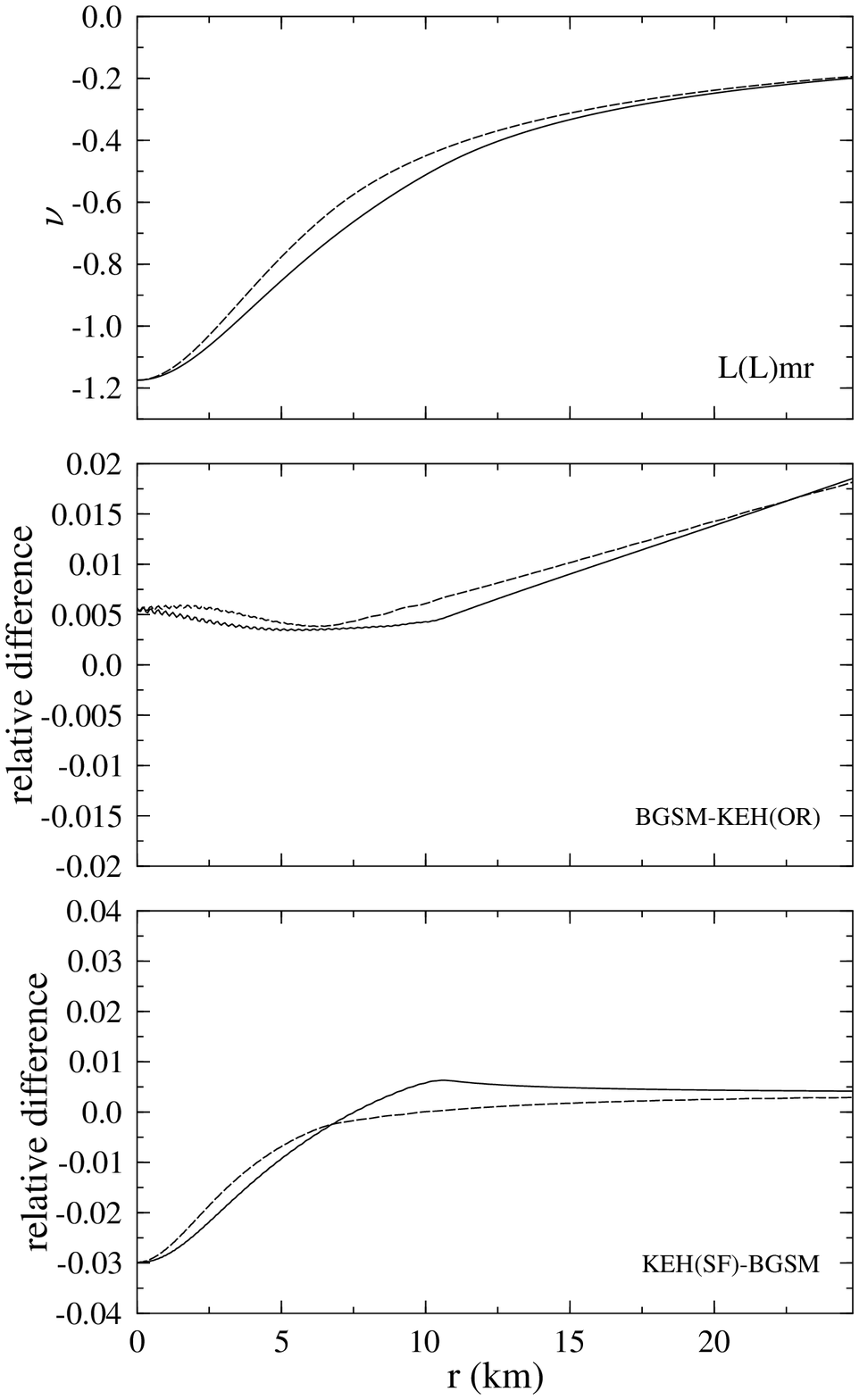}}
\caption{Same as Figure 1 but for the metric potential $\nu$ of model L(L)$mr$.}
\end{figure}

\begin{figure}
\resizebox{\hsize}{10.5cm}{\includegraphics{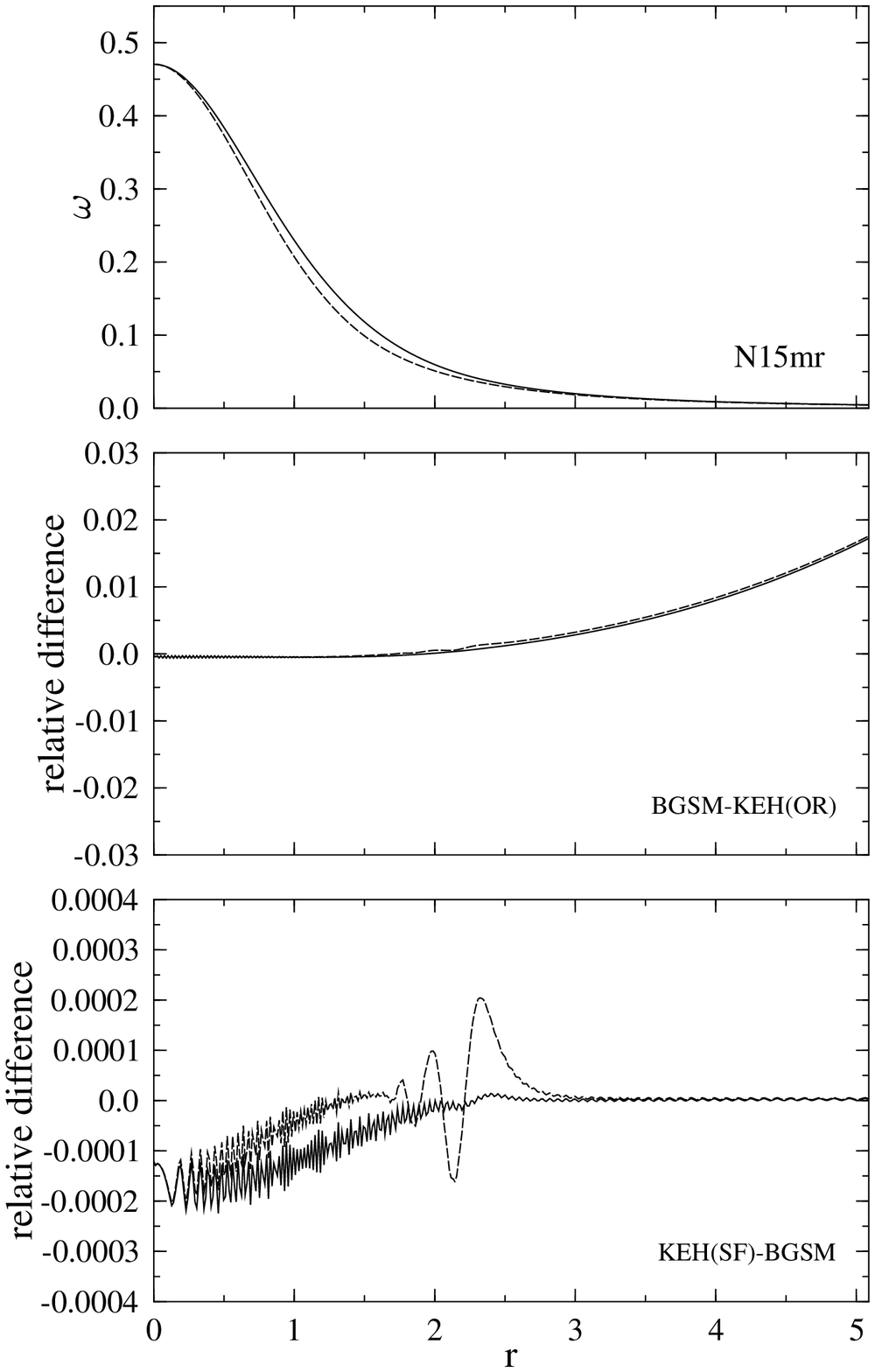}}
\caption{Same as Figure 1 but for the metric potential $\omega$ (in
units of $\Omega$) of model N15$mr$.}
\end{figure}

\begin{figure}
\resizebox{\hsize}{10.5cm}{\includegraphics{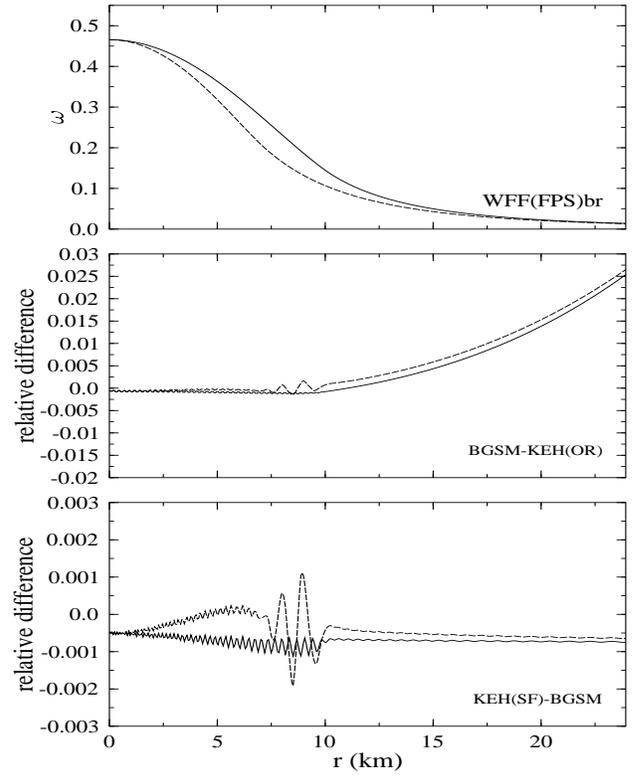}}
\caption{Same as Figure 1 but for the metric potential $\omega$ (in
units of $\Omega$) of model WFF(FPS)$br$.}
\end{figure}

\begin{figure}
\resizebox{\hsize}{10.5cm}{\includegraphics{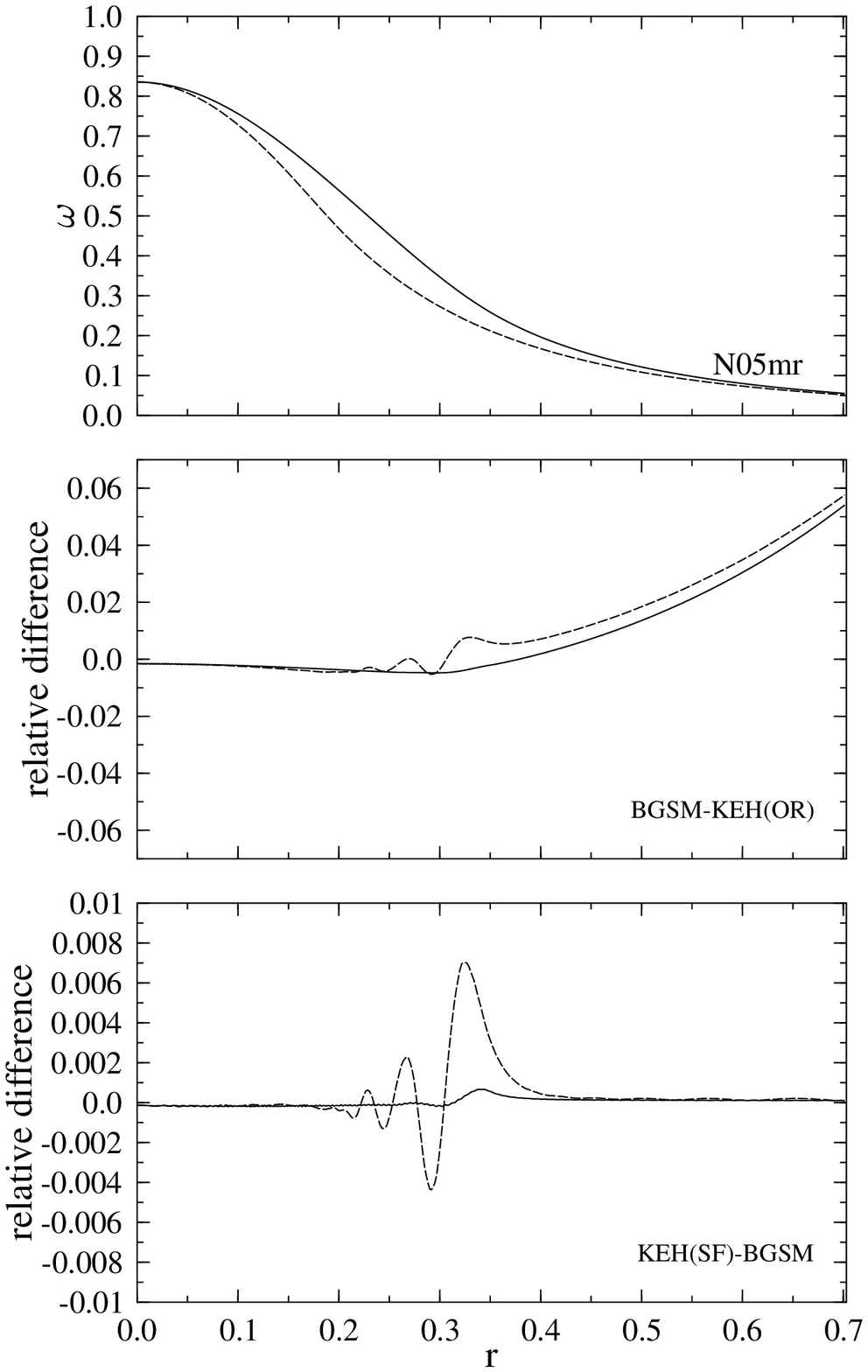}}
\caption{Same as Figure 1 but for the metric potential $\omega$ (in
units of $\Omega$) of model N05$mr$.}
\end{figure}

\begin{figure}
\resizebox{\hsize}{10.5cm}{\includegraphics{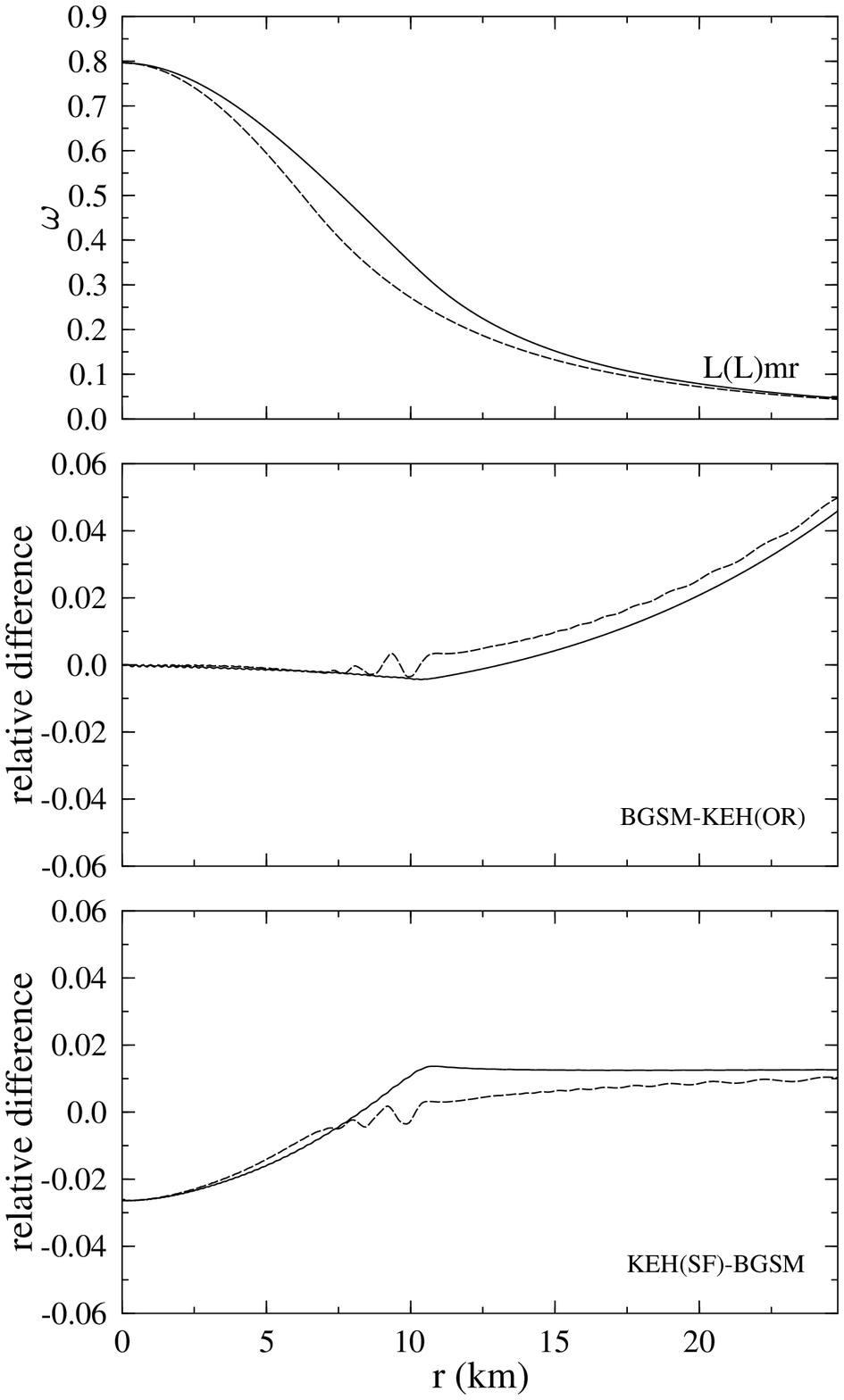}}
\caption{Same as Figure 1 but for the metric potential $\omega$ (in
units of $\Omega$) of model L(L)$mr$.}
\end{figure}

\section{Discussion and Conclusion}

\subsection{Discussion}

\subsubsection{Metric Potentials}

As redshift factors differ by about 10 \%  for the constant density, relativistic 
model N00$rr$ between the three codes (while the agreement global quantities 
is within a few  \%) we compare directly the local values of metric potentials for 
several models.  Figures 1 to 16 show the four metric potentials (upper panel)
and the relative differences in them between BGSM and KEH(OR) (middle panel)
and KEH(SF) and BGSM (lower panel) for the  models N05$mr$, N15$mr$, L(L)$mr$ 
and WFF(FPS)$br$. The metric potentials are graphed against the coordinate $r$ 
in the equatorial plane ($\theta= \pi/2$, solid line) and along the axis of rotation 
($\theta=0$, dashed line). The range of the coordinate $r$ is the twice the equatorial
radius of the star.

In general, the agreement in the local values of the metric potentials reflects the
agreement in the computed physical parameters of models. In these graphs, several
significant behaviors can be pointed out: First, there
are high frequency and small amplitude oscillations at the inner
part of the stars for all models. Second, the differences between the 
results of KEH(OR) and those of the other two codes are growing outside the stars
as $r$ increases. Third, although the differences between the KEH(SF) and BGSM codes
are very small for models N05$mr$, N15$mr$ and WFF(FPS)$br$, there appear larger
differences for the stiff  model L(L)$mr$. Fourth, there appears a larger amplitude
oscillation in the metric potential  $\omega$ on the axis of rotation, close to the
surface. 

The first behavior is due to the integration scheme of the KEH code,
i.e. the Simpson scheme. In general, the Simpson scheme gives results
with higher precision, compared with those obtained by the trapezoidal
scheme.  However, in the KEH scheme, the integrands contain nonsmooth
functions with respect to the radial coordinate, because of the nature
of the Green's functions. The non-uniform distribution of the weight
factor in Simpson's scheme for nonsmooth functions results in
oscillating behaviors with very small amplitudes, which cannot be
noticed in the behavior the integrated quantities.

The second behavior in the original KEH code is caused by the "truncation" 
of the domain of integration at a finite distance from the star, instead of 
integrating over the whole space.

The large differences in the metric potentials between KEH(SF) and BGSM 
for EOS L, could be accounted to the stiffness of the equation of state, but
the differences between KEH(OR) and BGSM for the same model are not as
large, and we have not an explanation for that.  

The oscillations in $\omega$ on the axis of rotation near the surface are present
also for the soft $N=1.5$ polytropes, while for $N=0.5$ they are larger. This
indicates that terms in the field equations for $\omega$ are very sensitive to
the presence of the surface and the accompanying Gibbs phenomenon. Even 
for $N=1.5$ polytropes, where the density goes to zero smoothly at the surface,
there is a small scale Gibbs phenomenon, due to the finite number of grid
points used to represent the region of integration.

\subsubsection{Method of Interpolation}

An important factor for the
local accuracy of models constructed with realistic equations of state
is the method of interpolation of the energy vs. pressure data given in
an EOS  table. While global quantities are not affected significantly, 
the virial identities for realistic EOSs, are sensitive to the interpolation scheme 
This can be considered to reflect the nature of the interpolation
scheme as mentioned before. If we define the enthalpy ($H$) by
\begin{equation}
H \equiv \ln \left( { \varepsilon + p \over \rho c^2} \right),
\end{equation}
the Gibbs-Duhem relation, which follows directly from the first law 
of thermodynamics, implies
\begin{equation}
        {dp\over dH} = \varepsilon + p  \ .
\end{equation}
In the cubic Hermite interpolation, the Gibbs-Duhem relation is used to
replace by $\nabla H$ the term $\nabla p /(\varepsilon+p)$ which appears 
in the hydrostationary equilibrium equation. If the tabulated function $p(H)$ fails to 
satisfy the above relation, then the hydrostationary equilibrium equation, which is
derived from the Bianchi identity, is only approximately verified 
by the numerical solution, which  results in increased error in the GRV2 and GRV3 
virial identities. 

The four point Lagrange interpolation does not satisfy the Gibbs-Duhem 
relation because it only reproduces the values of the discrete points,
but there is no guarantee for the reproduction of the derivatives. 
This explains why the GRV2 and GRV3 errors are bad, even in the
nonrotating case (GRV2 = 3E-03, GRV3 = 1E-02 for model L$sr$)
as compared to  ${\rm GRV2}\sim 10^{-14}$ for polytropic models (see e.g. 
Bonazzola et al. \cite{bona93}). The GRV2 and GRV3 error indicators thus do not reflect 
the precision of the code but the bad thermodynamical behavior of the 
tabulated EOS. 

The advantage of the cubic Hermite interpolation is that the
Gibbs-Duhem relation is automatically satisfied because this
interpolation reproduces not only the values themselves but also the
derivatives (Swesty 1996).
Moreover, in our case, the energy density and the baryon number density 
are obtained by
\begin{eqnarray}
        \varepsilon & = & {p \over H} {d\log p\over d\log H} - p,  \\
        n & = & {\varepsilon + p \over m_{\rm B} c^2} \exp(-H) \ .
\end{eqnarray}
Because of these equations, the Gibbs-Duhem relation is satisfied in every
point. Note also that the value of $\varepsilon$ obtained in this way coincides
exactly with $\varepsilon_i$ at the points in the tabulated equation of state.

\subsection{Conclusion}

The comparison of three different codes for constructing rapidly
rotating relativistic neutron star models demonstrates that the BGSM
and KEH schemes used are highly accurate for typical polytropic models
- when the field equations are solved to infinity - and for models
constructed with realistic equations of state, that do not have phase
transitions.  If one approximates neutron stars as constant density
stars, then Gibbs phenomena at the discontinuous surface reduce the
accuracy of the computed models.  If high accuracy in such models and
in models with phase transitions is desired, then modified numerical
schemes - free of Gibbs phenomena - need to be used.  Such numerical
schemes could employ, for example, surface fitted coordinates. Such a
scheme has been presented recently by Bonazzola et al. (1998a) in the
framework of spectral methods and looks promising for rotating stellar
models.  Further, we demonstrated that the metric potentials are
subject to various local oscillatory behaviors, even if integrated
quantities have very good accuracy.  This observation is important for
the effort of constructing initial data for the numerical evolution of
rotating relativistic neutron star models.

\begin{figure}
\resizebox{\hsize}{10.5cm}{\includegraphics{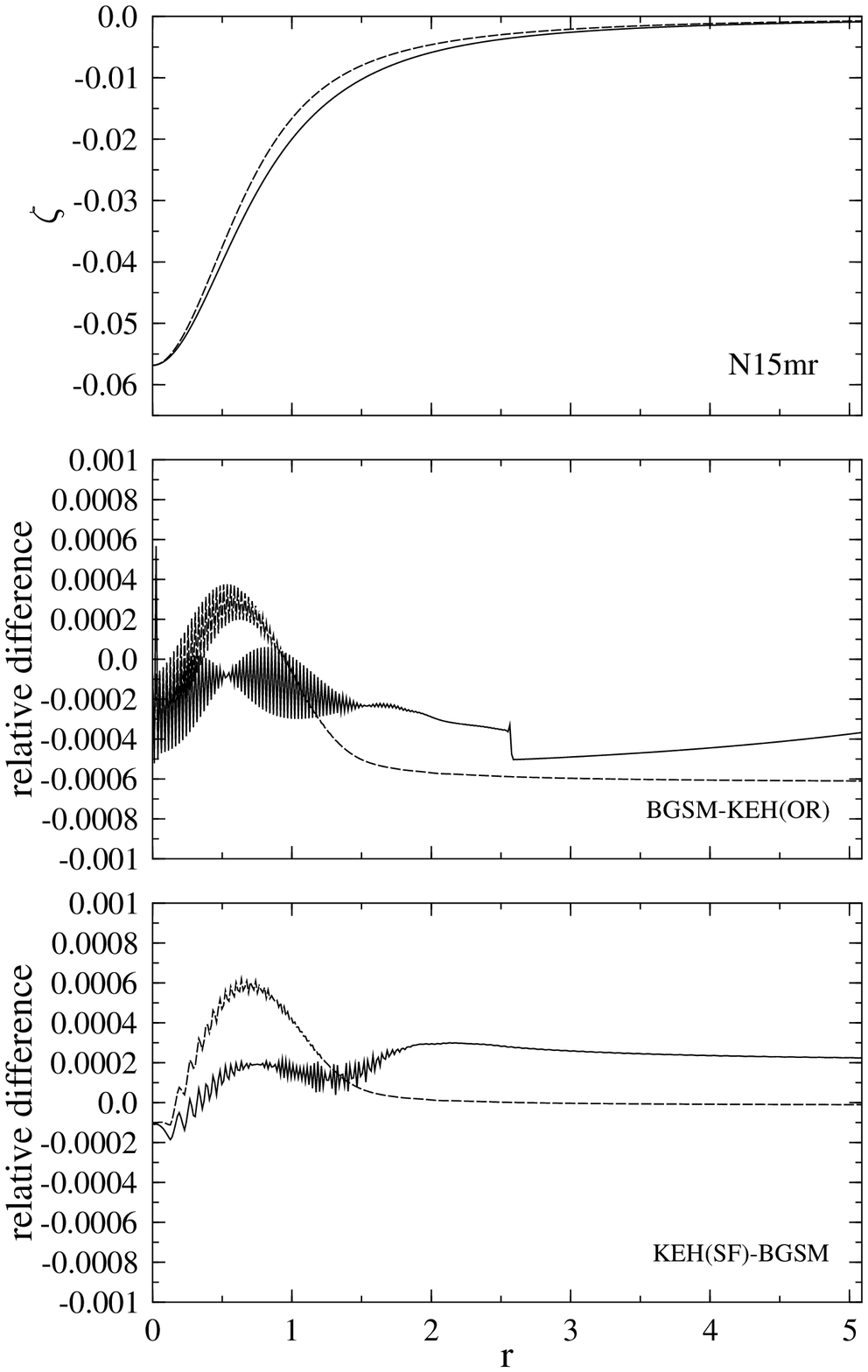}}
\caption{Same as Figure 1 but for the metric potential $\zeta$ of model N15$mr$.}
\end{figure}

\begin{figure}
\resizebox{\hsize}{10.5cm}{\includegraphics{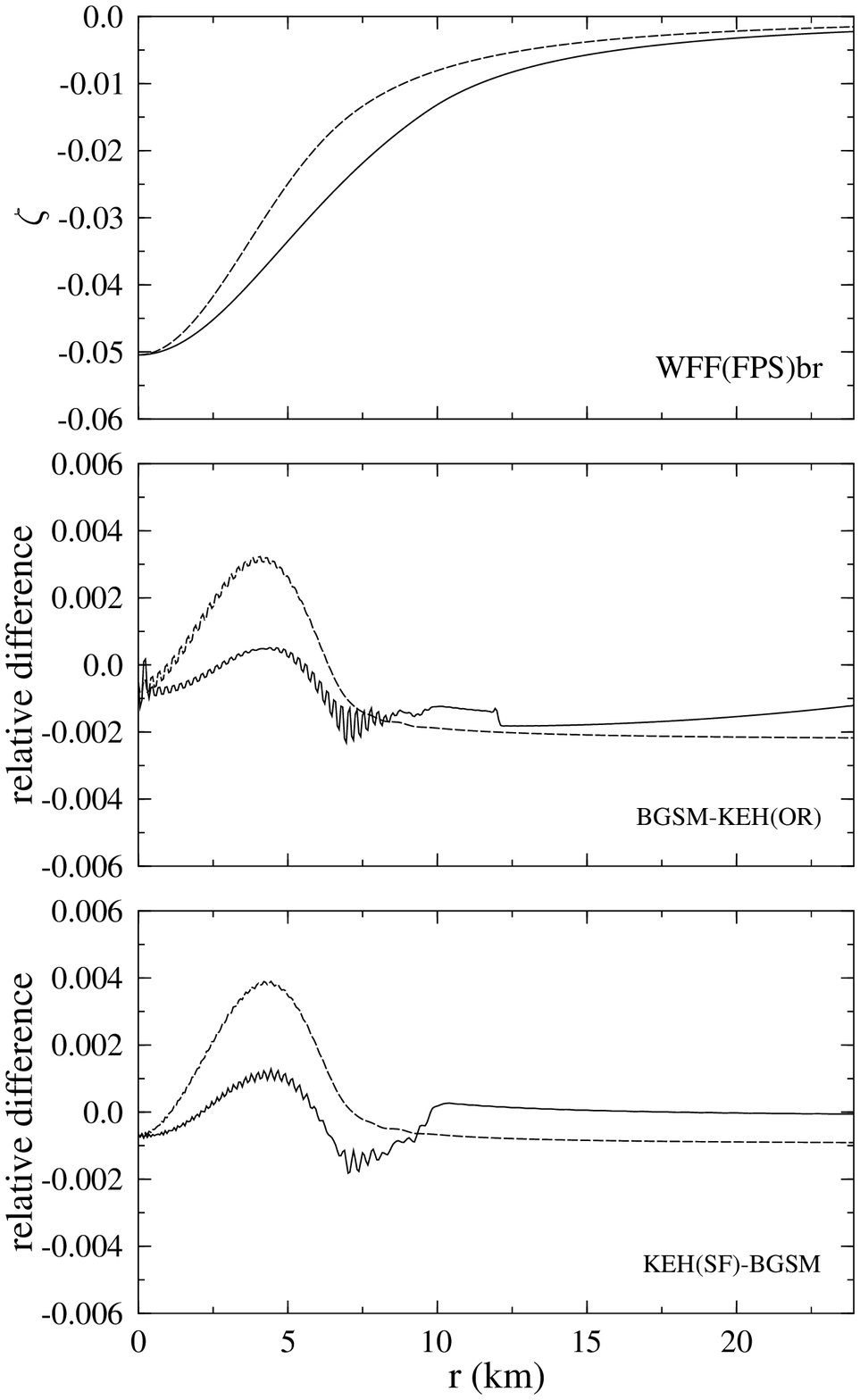}}
\caption{Same as Figure 1 but for the metric potential $\zeta$ of model WFF(FPS)$br$.}
\end{figure}

\begin{figure}
\resizebox{\hsize}{10.5cm}{\includegraphics{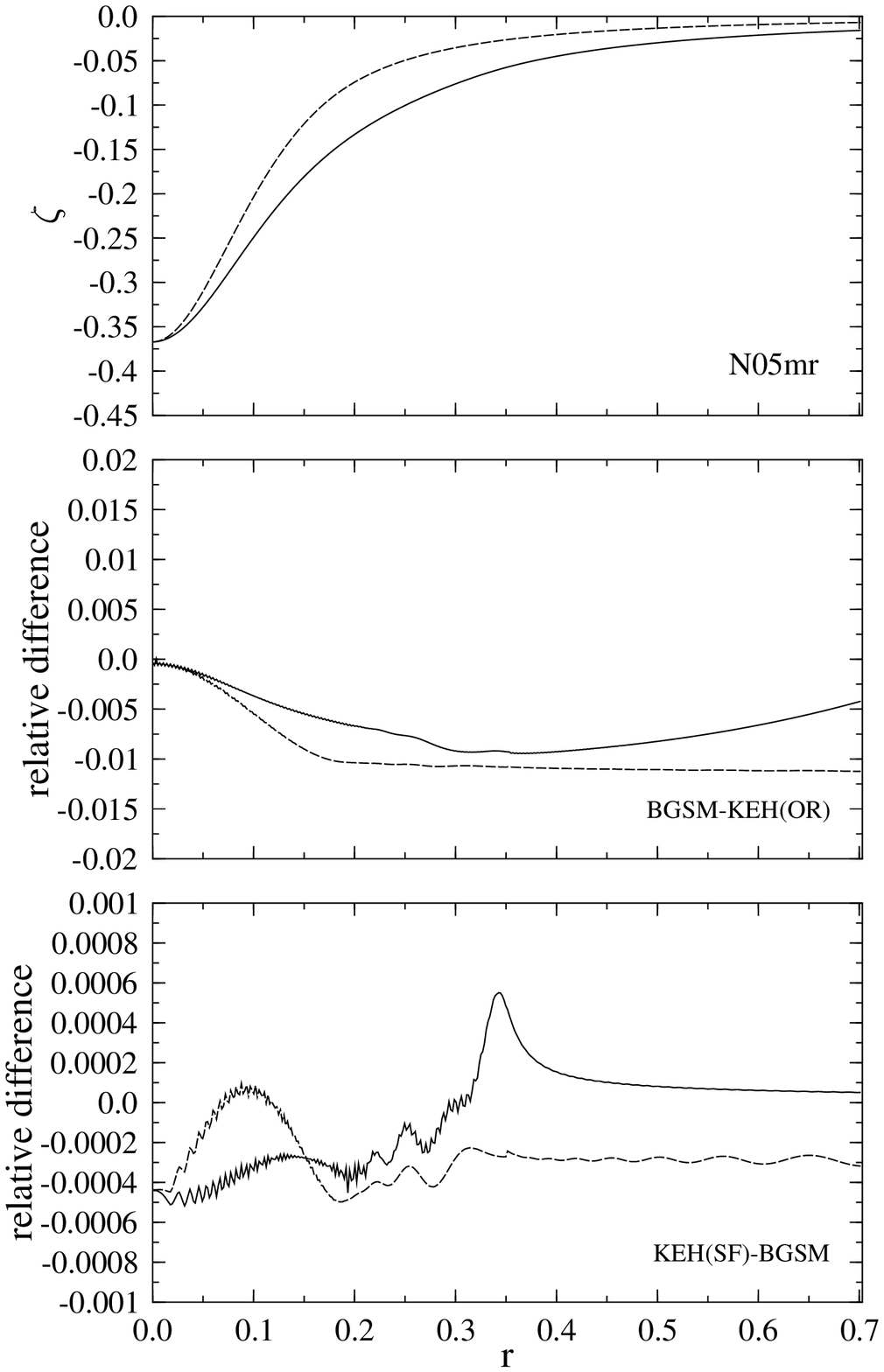}}
\caption{Same as Figure 1 but for the metric potential $\zeta$ of model N05$mr$.}
\end{figure}

\begin{figure}
\resizebox{\hsize}{10.5cm}{\includegraphics{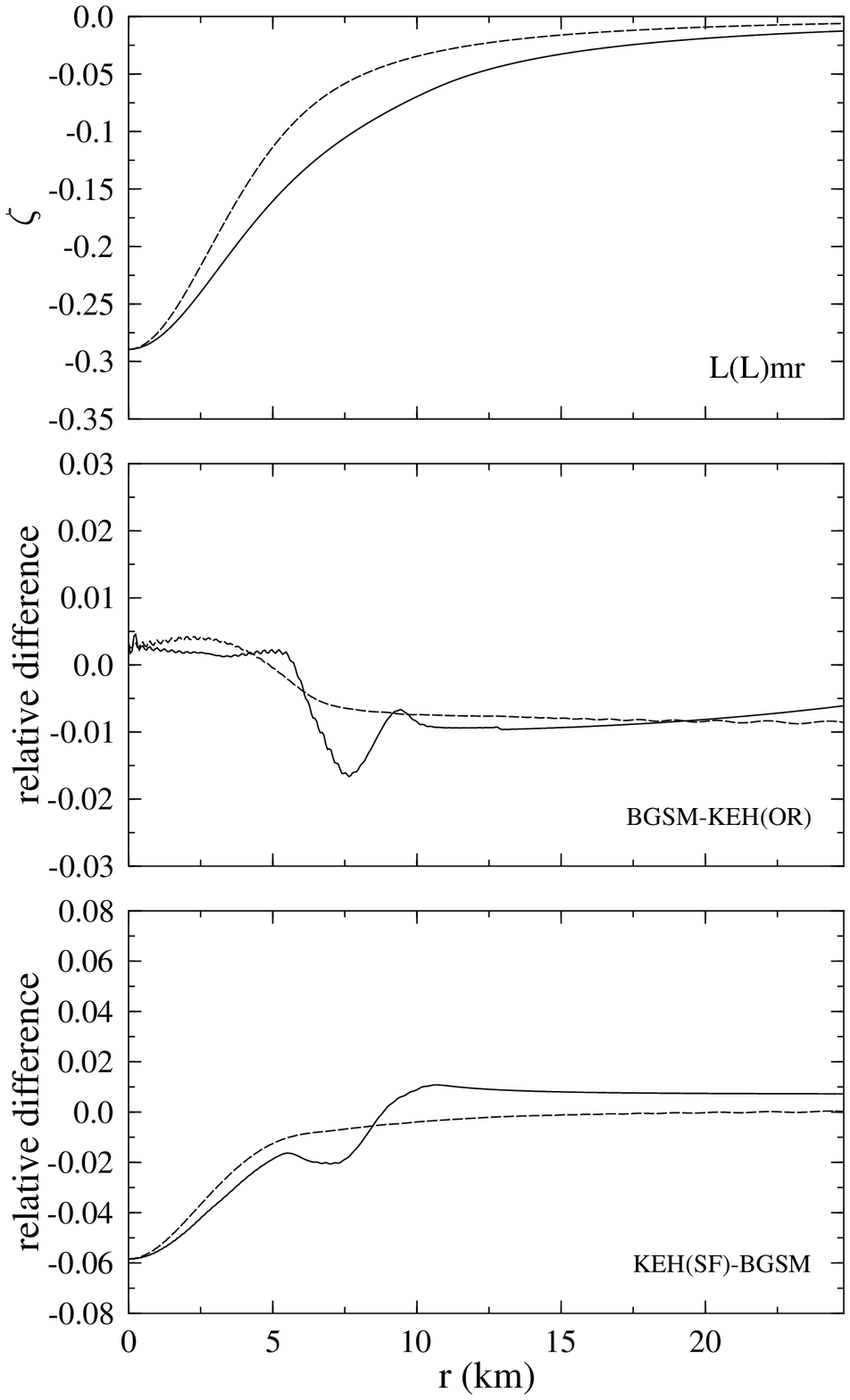}}
\caption{Same as Figure 1 but for the metric potential $\zeta$ of model L(L)$mr$.}
\end{figure}

\begin{acknowledgements}

We would like to thank S. Bonazzola and J. L. Friedman for very helpful 
discussions. This research has been supported 
%%%%%%%%%%%%%%%%%%%
in part
%%%%%%%%%%%%%%%%%%%
by NSF grant PHY-9507740
%%%%%%%%%%%%%%%%%%%
and by the Grant-in-Aid for Scientific Research of the Japanese Ministry
of Education, Science and Culture.
%%%%%%%%%%%%%%%%%%% 
N.S. wishes to acknowledge the generous hospitality 
of the Max Planck Institute for Gravitational Physics, Potsdam,  where part of this paper 
was completed.

\end{acknowledgements}


\begin{thebibliography}{}
\bibitem[1977]{arne77} Arnett, W.D., 
   \& Bowers, R.L. 1977, ApJS, 33, 415
\bibitem[1971a]{baym71a} Baym, G., Bethe, H.A. \& Pethick, C.J. 
   1971,  Nucl. Phys. A, 175, 225
\bibitem[1971b]{baym71b} Baym, G., Pethick, C.J. \& Sutherland, P.
   1971,  ApJ, 170, 299
\bibitem[1974]{beth74} Bethe, H.A. \& Johnson, M. 
   1974,  Nucl. Phys. A, 230, 1
\bibitem[1973]{bona73} Bonazzola, S. 1973,  ApJ, 182, 335
\bibitem[1996]{bona96}  Bonazzola, S., Frieben, J., 
   Gourgoulhon, E. \& Marck, J.A. 1996, in Proc. Third International
   Conference on Spectral and High Order Methods (ICOSAHOM'95), Eds.
   A.V.~Ilin \& L.R. Scott,
   (Houston Journal of Mathematics, Houston) 
\bibitem[1994]{bona94} Bonazzola, S. 
   \& Gourgoulhon, E. 1994, Class. Quantum. Grav., 11, 1775
\bibitem[1997]{bona97} Bonazzola, S., 
   Gourgoulhon, E., \& Marck, J.A. 1997, in Relativistic  Gravitation and
   Gravitational Radiation, Eds. J.-A.~Marck \& J.-P.~Lasota (Cambridge
   University Press, Cambridge)
\bibitem[1998a]{bona98a} Bonazzola, S., 
   Gourgoulhon, E., \& Marck, J.A. 1998a, Numerical approach for high precision 
   3-D relativistic star models,
   preprint astro-ph/9803086   
\bibitem[1998b]{bona98b} Bonazzola, S., 
   Gourgoulhon, E., \& Marck, J.A. 1998b, Spectral methods in relativistic
astrophysics, to appear in J. Comp. Appl. Math.   
\bibitem[1993]{bona93} Bonazzola, S.,  Gourgoulhon, E., 
   Salgado, M., \& Marck, J.A. 1993,  A\&A 278, 421
\bibitem[1974]{bona74}  Bonazzola, S. \&  Schneider,
   J. 1974,  ApJ, 191, 273 
\bibitem[1976]{butt76b} Butterworth, E.M. 1976,  ApJ, 204, 561
\bibitem[1975]{butt75} Butterworth, E.M. 
   \& Ipser, J.R. 1975,  ApJl, 200, L103
\bibitem[1976]{butt76a} Butterworth, E.M. 
   \& Ipser, J.R. 1976,  ApJ, 204, 200
\bibitem[1974]{canu74} Canuto, V. \& Chitre, S.M. 
   1974,  Phys. Rev. D, 9, 1587
\bibitem[1969]{car69} Carter, B. 1969, J. Math. Phys. 10, 70 
\bibitem[1992]{cook92} Cook, G.B., Shapiro, S.L. 
\& Teukolsky, S.A. 1992,  ApJ, 398, 203
\bibitem[1994a]{cook94a} Cook, G.B., Shapiro, S.L. 
   \& Teukolsky, S.A. 1994a,  ApJ, 422, 273
\bibitem[1994b]{cook94b} Cook, G.B., Shapiro, S.L. 
   \& Teukolsky, S.A. 1994b,  ApJ, 424, 823
\bibitem[1994]{erig94} Eriguchi, Y., Hachisu, I. 
   \& Nomoto, K. 1994,  MNRAS, 266, 179
\bibitem[1949]{feyn49} Feynman, R.P., Metropolis, N. \& 
   Teller, E. 1949, Phys. Rev., 75, 1561
\bibitem[1988]{frie88} Friedman, J.L., Imamura, J.N., 
   Durisen, R.H. \& Parker, L. 1988, Nature, 336, 560
\bibitem[1984]{frie84} Friedman, J.L., Ipser, J.R. 
   \& Parker, L. 1984, Nature, 312, 255
\bibitem[1986]{frie86} Friedman, J.L., Ipser, J.R. 
   \& Parker, L. 1986,  ApJ, 304, 115
\bibitem[1989]{frie89} Friedman, J.L., Ipser, J.R. 
   \& Parker, L. 1989, Phys. Rev. Lett., 62, 3015
\bibitem[1981]{frie81} Friedman, B. 
   \& Pandharipande, V.R. 1981,  Nucl. Phys. A, 361, 502 
\bibitem[1994]{gour94} Gourgoulhon, E. 
   \& Bonazzola, S. 1994, Class. Quantum. Grav., 11, 443
\bibitem[1977]{gott77}  Gottlieb, D. \& Orszag, S.
   1977, Numerical Analysis of Spectral Methods: Theory and Application,
   Regional Conference Series in Applied Mathematics, Vol. 26
   (Philadelphia: SIAM) 
\bibitem[1986]{hach86} Hachisu, I. 1986,  ApJS, 61, 479
\bibitem[1989a]{koma89a} Komatsu, H., Eriguchi 
   \& Hachisu, I. 1989a,  MNRAS, 237, 355
\bibitem[1989b]{koma89b} Komatsu, H., Eriguchi 
   \& Hachisu, I. 1989b,  MNRAS, 239, 153
\bibitem[1993]{lore93} Lorenz, C.P., Ravenhall, D.G. 
   \& Pethick, C.J. 1993, Phys. Rev. Lett., 70, 379
\bibitem[1973]{nege73} Negele, J.W. \& Vautherin, D. 
   1973,  Nucl. Phys. A, 207, 298 
\bibitem[1968]{ostr68} Ostriker, J.P. \& Mark, J.W.-K. 
   1968,  ApJ, 151, 1075
\bibitem[1971]{pand71} Pandharipande, V.R. 1971, 
    Nucl. Phys. A, 178, 123 
\bibitem[1976]{pand76} Pandharipande, V.R. 
   Pines, D. \& Smith, R.A. 1976,  ApJ, 208, 550 
\bibitem[1975]{pand75} Pandharipande, V.R. 
   \& Smith, R.A. 1975, Phys. Lett., 59B, 15 
\bibitem1966]{papa66} Papapetrou, A. 1966, Ann. Inst. H.
   Poincar\'e, A4, 83 
\bibitem[1994a]{salg94} Salgado, M, Bonazzola, S., 
   Gourgoulhon, E. \& Haensel, P. 1994a,  A\&A, 291, 155
\bibitem[1994b]{salg94b}  Salgado, M, Bonazzola, S., 
   Gourgoulhon, E. \& Haensel, P. 1994b,  A\&As, 108, 455 
\bibitem[1995]{stag95} Stergioulas, N. 
  \& Friedman, J.L. 1995,  ApJ, 444, 306
\bibitem[1998]{S98} Stergioulas, N. 1998, ``Rotating Stars in Relativity'',
to appear in {\rm Living Reviews in Relativity}, http://www.livingreviews.org 
\bibitem[1996]{swes96} Swesty, F.D. 1996, 
 J. Comp. Phys., 127, 118
\bibitem[1978]{tass78} Tassoul, J.L. 1978, Theory of Rotating
  Stars (Princeton Univ. Press : Princeton, NJ)
\bibitem[1989]{tayl89} Taylor, J.H. 
  \& Weisberg, J.M. 1989,  ApJ, 345, 434
\bibitem[1965]{toop65} Tooper, R.F. 1965,  ApJ, 142, 1541
\bibitem[1993]{thor93} Thorsett, S.E.,
  Arzoumanian, Z., McKinnon, M.M. \&  Taylor, J.H. 1993,  ApJ, 405, L29
\bibitem[1995]{vankerk95}  van Kerkwijk, M.H.,
  van Paradijs, J. \&  Zuiderwijk, E.J 1995,  A\&A, 303, 497 
\bibitem[1988]{wiri88} Wiringa, R.B., Fiks, V. 
  \& Farbrocini, A. 1988, Phys. Rev. C, 38, 1010
\bibitem[1997]{wol97} Wolszczan, A. 1997, 
        in Relativistic  Gravitation and
   Gravitational Radiation, Eds. J.-A.~Marck \& J.-P.~Lasota (Cambridge
   University Press, Cambridge)
\end{thebibliography}
\end{document}